\documentclass[11pt]{article}

\usepackage{
    amssymb,
    amsmath,
    amsfonts,
    amsthm,
    calc,
    eurosym,
    geometry,
    ulem,
    graphicx,
    caption,
    color,
    xcolor,
    setspace,
    sectsty,
    comment,
    footmisc,
    caption,
    pdflscape,
    subcaption,
    subfiles,
    titling,
    array,
    booktabs,
    longtable,
    float,
    authblk,
    makecell,
    bbm,
    threeparttable, 
    minitoc,
    tocbasic,
    enumitem}

\usepackage[page]{appendix} % print appendices title

\definecolor{forestgreen}{RGB}{34,139,34}
\definecolor{bblue}{HTML}{1E88E5} 
\definecolor{bgreen}{HTML}{004D40}
 \definecolor{bred}{HTML}{D81B60}
\usepackage{hyperref}
\hypersetup{
    colorlinks=true,
    linkcolor=magenta,
    filecolor=magenta, 
    citecolor=forestgreen,      
    urlcolor=blue
}

\DeclareTOCStyleEntry[dynnumwidth=true,
                      numsep=1em,
]{tocline}{section}
\DeclareTOCStyleEntry[dynnumwidth=true,
                      numsep=1em,
]{tocline}{subsection}

\usepackage[
    backend=biber,
    style=nature,
    date=year,
    doi=true,
    isbn=false,
    url=false,
    eprint=false
]{biblatex}

\AtEveryBibitem{%
  \clearfield{note}%
}
\AtEveryCitekey{\clearlist{publisher}}
\AtEveryBibitem{\clearlist{publisher}}

\usepackage{pgf,tikz}
\usetikzlibrary{arrows,automata}
\usetikzlibrary{shapes.geometric,positioning}
\usetikzlibrary{positioning,calc}
\usetikzlibrary{arrows,shapes.arrows,shapes.geometric,
shapes.multipart,decorations.pathmorphing,positioning,
swigs}
\usepackage{siunitx}
\newcolumntype{d}{S[table-format = 1.3]}

\newtheorem{proposition}{Proposition}

\theoremstyle{definition}

\newtheorem*{remark}{Remark}

\newcolumntype{L}[1]{>{\raggedright\arraybackslash}m{#1}}
\newcolumntype{C}[1]{>{\centering\arraybackslash}m{#1}}
\newcolumntype{R}[1]{>{\raggedleft\arraybackslash}m{#1}}
\subsubsectionfont{\normalfont\itshape}

\usepackage{mathtools}

\usepackage{letltxmacro}
\LetLtxMacro\orgvdots\vdots
\LetLtxMacro\orgddots\ddots

\makeatletter
\DeclareRobustCommand\vdots{%
  \mathpalette\@vdots{}%
}
\newcommand*{\@vdots}[2]{%
  \sbox0{$#1\cdotp\cdotp\cdotp\m@th$}%
  \sbox2{$#1.\m@th$}%
  \vbox{%
    \dimen@=\wd0 %
    \advance\dimen@ -3\ht2 %
    \kern.5\dimen@
    \dimen@=\wd2 %
    \advance\dimen@ -\ht2 %
    \dimen2=\wd0 %
    \advance\dimen2 -\dimen@
    \vbox to \dimen2{%
      \offinterlineskip
      \copy2 \vfill\copy2 \vfill\copy2 %
    }%
  }%
}
\DeclareRobustCommand\ddots{%
  \mathinner{%
    \mathpalette\@ddots{}%
    \mkern\thinmuskip
  }%
}
\newcommand*{\@ddots}[2]{%
  \sbox0{$#1\cdotp\cdotp\cdotp\m@th$}%
  \sbox2{$#1.\m@th$}%
  \vbox{%
    \dimen@=\wd0 %
    \advance\dimen@ -3\ht2 %
    \kern.5\dimen@
    \dimen@=\wd2 %
    \advance\dimen@ -\ht2 %
    \dimen2=\wd0 %
    \advance\dimen2 -\dimen@
    \vbox to \dimen2{%
      \offinterlineskip
      \hbox{$#1\mathpunct{.}\m@th$}%
      \vfill
      \hbox{$#1\mathpunct{\kern\wd2}\mathpunct{.}\m@th$}%
      \vfill
      \hbox{$#1\mathpunct{\kern\wd2}\mathpunct{\kern\wd2}\mathpunct{.}\m@th$}%
    }%
  }%
}
\newcommand{\vast}{\bBigg@{4}}
\newcommand{\Vast}{\bBigg@{5}}
\makeatother

\newcommand*\circled[1]{\tikz[baseline=(char.base)]{
            \node[shape=circle,draw,inner sep=2pt] (char) {#1};}}

\DeclareMathOperator{\E}{E}
\DeclareMathOperator{\indep}{\perp\!\!\!\perp}
\def\expit{\mathrm{expit}}

\begin{document}
%TC:ignore
\begin{titlepage}
\title{Identification and estimation of vaccine effectiveness in the test-negative design under equi-confounding}
\author[1,2]{Christopher B. Boyer\thanks{Corresponding author, email: \href{mailto:boyerc5@ccf.org}{boyerc5@ccf.org}}}
\author[3]{Kendrick Qijun Li}
\author[4]{Xu Shi}
\author[5]{Eric J. Tchetgen Tchetgen}
\affil[1]{Department of Quantitative Health Sciences, Cleveland Clinic, Cleveland, OH.}
\affil[2]{Department of Medicine, Cleveland Clinic Lerner College of Medicine of Case Western Reserve University, Cleveland, OH.}
\affil[3]{Department of Biostatistics, St. Jude Children's Research Hospital, Memphis, TN.}
\affil[4]{Department of Biostatistics, University of Michigan, Ann Arbor, MI.}
\affil[5]{Department of Statistics and Data Science, The Wharton School, University of Pennsylvania, Philadelphia, PA.}

\date{\today\vspace{-1em}}
\maketitle

\begin{abstract}
The test-negative design (TND) is widely used to evaluate vaccine effectiveness in real-world settings. In a TND study, individuals with similar symptoms who seek care are tested, and effectiveness is estimated by comparing vaccination histories of test-positive cases and test-negative controls. The TND is often justified on the grounds that it reduces confounding due to unmeasured health-seeking behavior, although this has not been formally described using potential outcomes. At the same time, concerns persist that conditioning on test receipt can introduce selection bias. We provide a formal justification of the TND under an assumption of odds ratio equi-confounding, where unmeasured confounders affect test-positive and test-negative individuals equivalently on the odds ratio scale. Health-seeking behavior is one plausible example. We also show that these results hold under the outcome-dependent sampling used in TNDs. We discuss the design implications of the equi-confounding assumption and provide alternative estimators for the marginal risk ratio among the vaccinated under equi-confounding, including outcome modeling and inverse probability weighting estimators as well as a semiparametric estimator that is doubly robust.  When equi-confounding does not hold, we suggest a straightforward sensitivity analysis that parameterizes the magnitude of the deviation on the odds ratio scale. A simulation study evaluates the empirical performance of our proposed estimators under a wide range of scenarios. Finally, we discuss broader uses of test-negative outcomes to de-bias cohort studies in which testing is triggered by symptoms.

\bigskip
\end{abstract}
\setcounter{page}{0}
\thispagestyle{empty}
\end{titlepage}
\pagebreak \newpage
%TC:endignore

%\doublespacing
\doparttoc % Tell to minitoc to generate a toc for the parts
\faketableofcontents % Run a fake tableofcontents command for the partocs

\part{} % Start the document part

\section*{Introduction} \label{sec:introduction}
Post-market evaluation of vaccine effectiveness is critical for understanding how vaccines perform in real-world settings, especially as population immunity changes or pathogens undergo antigenic mutation or selection \cite{patel_postlicensure_2020,hitchings_effectiveness_2021,israel_elapsed_2021}. These studies can also assess effectiveness in subpopulations that were ineligible or underrepresented \cite{olson_effectiveness_2022}, evaluate outcomes and safety signals that may have been underpowered \cite{thompson_effectiveness_2021}, or compare vaccine formulations not considered in the original trials \cite{skowronski_two-dose_2022}. Often, these evaluations must rely on observational data \cite{chua_use_2020-1,dean_covid-19_2021}. 

The test-negative design (TND) is a common and cost-efficient study design for estimating vaccine effectiveness \cite{sullivan_potential_2014,jackson_test-negative_2013}. In its canonical form, patients with acute respiratory illness are prospectively screened and recruited based on a unified symptom set through hospitals or other care facilities and tested for the pathogen of interest. Vaccination history and other detailed covariate information are collected at enrollment or through linked medical or vaccination records. Effectiveness is then estimated by comparing the odds of vaccination among test-positive (cases) and test-negative (controls) individuals.

The TND has been previously promoted as a way to reduce confounding due to unmeasured healthcare-seeking behavior  --- a common source of bias in observational evaluations of vaccines \cite{jackson_test-negative_2013}. Because participants seek a test conditional on the same symptoms but without knowledge of the causative agent, the design effectively restricts to ``care-seekers'', thereby precluding it as potential confounder \cite{jackson_test-negative_2013}. Subsequent work pointed out that the TND (i) may not be robust to unmeasured healthcare-seeking when it is nonbinary or when there are other factors affecting vaccine uptake and risk of infection \cite{sullivan_theoretical_2016,lewnard_theoretical_2021,lipsitch_observational_2016} and (ii) is subject to possible selection bias due to conditioning on receiving a test \cite{sullivan_theoretical_2016,lipsitch_observational_2016}.

More recently, \textcite{schnitzer_estimands_2022} and \textcite{jiang_tnddr_2023} introduced a formal causal framework for the TND based on potential outcomes, allowing for a more precise discussion of identification of marginal and conditional vaccine effects in the TND. However, contrary to the original motivation for the TND, they require that observed covariates are sufficient to ensure no unmeasured confounding or selection bias. Thus, a question remains as to whether the TND may be formally justified under weaker assumptions.

In this article, we revisit the original motivation and show that the TND may be justified under a less restrictive assumption that unmeasured confounding is equivalent for test-positive cases and test-negative controls on the odds ratio scale. This assumption builds on the insight that infection with another respiratory illness, if unaffected by focal vaccination, is essentially a negative outcome control \cite{lipsitch_negative_2010,shi_selective_2020,piccininni_using_2024}. Therefore, any observed difference in the incidence of test negative illnesses between vaccinated and unvaccinated individuals likely reflects residual confounding. Under the assumption of confounding equivalence, we can  use the difference in test-negative illness to de-bias our estimate of vaccine effectiveness, similar to the bias correction via the parallel trends assumption in the difference-in-differences literature \cite{sofer_negative_2016,park_universal_2023,tchetgen_universal_2023}. As we show, this approach is valid under the outcome-dependent sampling of the TND, where only those who are tested are sampled. When our assumptions hold, we show that the traditional odds ratio estimator identifies the conditional causal risk ratio \textit{among the vaccinated}. We also derive new estimators of the marginal risk ratio among the vaccinated, including those based on outcome modeling and inverse probability weighting. We discuss the implications of our assumptions for the design of TNDs and provide proofs of our results as well as simulations to illustrate the consequences when our assumptions are violated. We also propose a sensitivity analysis to evaluate the impact of deviations from equi-confounding on our effect estimates. Finally, although our focus is on the TND, we show how our bias adjustment approach under equi-confounding applies more broadly to cohort studies with symptom-triggered testing, i.e., where the outcome is only ascertained for those who seek a test after developing symptoms.

\section*{Setup and observed data} \label{sec:setup}
Let $X$ represent a vector of baseline covariates, $V$ an indicator of vaccination status (1: vaccinated, 0: unvaccinated), and $Y$ a tri-level outcome at the end of follow up where:
$$Y:= \begin{cases}
    Y = 2 & \text{test positive case}, \\
    Y = 1 & \text{test negative case}, \\
    Y = 0 & \text{not tested}.
\end{cases}
$$
For simplicity, we consider the original TND setting in which a vaccination campaign occurs prior to a seasonal outbreak, and thus vaccination is a point treatment. 

The outcome $Y$ is only meaningful with additional context about the testing regime in the population of interest. We introduce two additional variables that jointly define this regime. First, a categorical indicator $I$ of symptomatic infection (2: caused by pathogen of interest, 1: other cause, 0: no symptoms), where symptoms are the pre-specified set used to screen for inclusion in the TND study. Second, an indicator of receiving a test at a health facility $T$ (1: tested, 0: not tested). 

We further assume:
\begin{enumerate}
    \item[(i)] The chance of co-infection or repeated infections of different origin are negligible over the follow up interval such that $I = 1$ and $I = 2$ are mutually exclusive events;
    \item[(ii)] individuals are screened for symptoms prior to testing such that $T_i = 1 \implies I_i \neq 0$ for every individual $i$;
    \item[(iii)] all screened individuals are tested;
    \item[(iv)] the test is perfect such that $Y_i = 2 \implies I_i = 2$ and $Y_i = 1 \implies I_i = 1$ for every individual $i$.
\end{enumerate}
These assumptions can be relaxed (e.g., we discuss imperfect testing in \ref{sec:testing}). Under this testing regime, we have that $Y = I \times T$  and we consider $Y \neq 0$ to define \textit{medically attended illness}, i.e., symptomatic infection severe enough to warrant medical attention. Finally, let $Y^v$, $I^v$ and $T^v$ denote the potential outcomes of $Y$, $I$ and $T$ respectively under $V=v$ for $v=0,1$. 

The observed TND data are $n$ independent realizations 
$$O_{\text{TND}} = \{(X_i, V_i, S_i = 1, Y_i)\}_{i=1}^n,$$
from the law of $(X, V, Y) \mid S = 1$ where $S$ denotes selection into the TND study with $S = \mathbbm 1(Y \neq 0)$ under the testing regime.

For comparison, we also define data from a hypothetical cohort, i.e. $N$ independent realizations 
$$O_{\text{cohort}} = \{(X_i, V_i, Y_i)\}_{i=1}^N,$$
from the law of $(X, V, Y)$, under the same screening and testing regime. TND data can be recovered by restricting the cohort to those tested, but not vice versa. 

\section*{Causal estimands for the test-negative design} \label{sec:estimands}
In a TND study, the primary estimand is the marginal causal risk ratio,
\begin{equation*}
    RR \equiv \dfrac{\Pr[Y^1 = 2]}{\Pr[Y^0 = 2]},
\end{equation*}
comparing the probability of medically attended illness due to the pathogen of interest if everyone were vaccinated versus unvaccinated in the population underlying the TND, with causal \textit{vaccine effectiveness} (VE) defined as $1 - RR$. Occasionally, the causal odds ratio is  presented as the target parameter instead of the risk ratio. However, this is often combined with an assumption ---either explicit or implicit--- that symptomatic infection is rare, in which case the odds ratio approximates the risk ratio. 

Under the assumption that the measured covariates $X$ suffice to control confounding and selection bias, \Citeauthor*{schnitzer_estimands_2022} \cite{schnitzer_estimands_2022} showed that $RR$ can be identified by the odds ratio comparing vaccination among test-positive cases and test-negative controls using TND data. When unconfoundedness does not hold, $RR$ is generally not identified. However, we show that an alternative estimand of interest, the causal risk ratio \textit{among the vaccinated}, i.e.
\begin{equation*}
    \Psi \equiv \dfrac{\Pr[Y^1=2 | V = 1]}{\Pr[Y^0 = 2 | V = 1]},
\end{equation*}
may nevertheless be point identified. The parameter $\Psi$ is analogous to the average treatment effect on the treated (ATT) in the causal inference literature. Similarly, we can define a modified VE as $1-\Psi$, interpreted as the proportion of medically attended illness prevented by vaccination among the vaccinated subjects. When protection is homogeneous, $\Psi$ coincides with the causal risk ratio, i.e. $RR = \Psi$; otherwise, when protection is heterogeneous, $RR$ is a weighted average of the causal risk ratios among the vaccinated and unvaccinated. 

We also consider the \textit{conditional} causal risk ratio among the vaccinated, i.e. 
\begin{equation*}
    \Psi(X) \equiv \frac{\Pr[Y^1 = 2 | V = 1, X]}{\Pr[Y^0 = 2 | V = 1, X]}.
\end{equation*}
which is the estimand targeted by traditional logistic regression estimator in a TND study when effect homogeneity is assumed or when heterogeneity is modeled via inclusion of product terms. While $\Psi$ is our preferred estimand, we use $\Psi(X)$ to first present our identification arguments in simplified form.

\section*{Illustrative example} \label{sec:roadmap}
To build intuition, we start with a brief account of the identification of the conditional risk ratio among the vaccinated, $\Psi(X)$, before turning to formal results for $\Psi$. 

\subsection*{Decomposition of the conditional causal risk ratio among the vaccinated}
Consider the estimand $\Psi(X)$. Multiplying by unity in the form of $\frac{\Pr[Y^0 = 2 | V = 0, X]}{\Pr[Y^0 = 2 | V = 0, X]}$ and applying consistency of potential outcomes (Assumption \ref{ass1} below), we can write:
\begin{align} \label{eqn:decomposition}
    \Psi(X) &=\underbrace{\frac{\Pr[Y = 2| V = 1, X]}{\Pr[Y = 2 | V = 0, X]}}_{\text{observed risk ratio}} \times \underbrace{\frac{\Pr[Y^0 = 2 | V = 0, X]}{\Pr[Y^0 = 2 | V = 1, X]}}_{\text{de-biasing term}}.
\end{align}
This decomposition expresses $\Psi(X)$ as the product of two terms: the (crude) observed risk ratio and a de-biasing term. The de-biasing term quantifies the degree of unmeasured confounding between the vaccinated and unvaccinated on the ratio scale, i.e., the net influence of all unmeasured causes of vaccination and the outcome. When there is no unmeasured confounding, the de-biasing term equals one and the observed risk ratio equals the causal risk ratio among the vaccinated. Conversely, in the presence of unmeasured confounding, the de-biasing term is the quantity necessary to remove bias from the observed risk ratio. However, without additional assumptions, this term (in particular the denominator) cannot be identified from the observed data. 

\subsection*{The test-negative illness as a proxy for unmeasured confounding}
To identify the de-biasing term, $\frac{\Pr[Y^0 = 2 | V = 0, X]}{\Pr[Y^0 = 2 | V = 1, X]}$, we seek a proxy outcome that could be used in place of test-positive illness for bias correction. A natural choice is another infection that produces the same symptoms and shares many of the same confounders, but is itself unaffected by focal vaccination. In a TND this is the role of the test-negative illness. When the net effect of all unmeasured factors $U$ is equivalent on the ratio scale for test-positive and test-negative illness after conditioning on covariates $X$, i.e. when
\begin{equation}\label{eqn:proxy}
     \dfrac{\Pr[Y^0 = 2  | V = 1, X]}{\Pr[Y^0 = 2 | V = 0, X]} = \frac{\Pr[Y^0 = 1 | V = 1, X]}{\Pr[Y^0 = 1 | V = 0, X]},
\end{equation}
we call this \textit{odds ratio equi-confounding} (Assumption \ref{ass3} below; Note in this case, because the illnesses are also assumed to be mutually exclusive, the odds ratio scale is also the risk ratio scale). Furthermore, if vaccination does not affect the test-negative outcome then 
\begin{equation}\label{eqn:no_effect}
     \frac{\Pr[Y^0 = 1 | V = 1, X]}{\Pr[Y^0 = 1 | V = 0, X]} = \frac{\Pr[Y = 1 | V = 1, X]}{\Pr[Y = 1 | V = 0, X]},
\end{equation}
implying that the de-biasing term is identified by the ratio of test-negative outcome comparing the vaccinated to unvaccinated.

\subsection*{Identification under cohort and outcome-dependent sampling}
Substituting $\frac{\Pr[Y = 1 | V = 1, X]}{\Pr[Y = 1 | V = 0, X]}$ for the de-biasing term in our original decomposition of $\Psi(X)$, we derive the following expression:
    \begin{equation}\label{eqn:or_estimand}
         \Psi_{om}(X) \equiv \dfrac{\Pr[Y = 2 | V = 1, X]/\Pr[Y = 1 | V = 1, X]}{\Pr[Y = 2 | V = 0, X]/\Pr[Y = 1 | V = 0, X]},
    \end{equation}
where the subscript \textit{om} is used to signify that $\Psi_{om}(X)$ is amenable to estimation via outcome modeling in contrast to inverse probability weighting estimands to be defined later. We note that $\Psi_{om}$ is the familiar difference-in-differences estimand on the multiplicative scale, with test-negative illness playing the role of the pre-treatment outcome --- essentially acting as a ``parallel'' trend for test-positive illness in the absence of vaccination. Importantly, $\Psi_{om}(X)$ is expressed solely in terms of observable quantities, meaning it could, in principle, be estimated using data from the underlying cohort. However, it is not identified under the sampling design of the TND which implicitly conditions on $S_i = 1$. 

Nevertheless, under the invariance of the odds ratio \cite{breslow_regression_1976}, we can show that $\Psi_{om}(X)$ is equivalent to
\begin{equation}\label{eqn:or_estimand_tnd}
    \Psi^*_{om}(X) = \dfrac{\Pr[Y = 2 | S = 1, V = 1, X]/\Pr[Y = 1 | S = 1, V = 1, X]}{\Pr[Y = 2 | S = 1, V = 0, X]/\Pr[Y = 1 | S = 1, V = 0, X]}
\end{equation}  
which is identified under the sampling design of the TND. As discussed further in section \ref{sec:estimation}, $\Psi^*_{om}(X)$ is identical to the conditional odds ratio estimand commonly targeted in TND studies \cite{jackson_test-negative_2013}, suggesting that estimates from existing studies could be reinterpreted and justified under the equi-confounding conditions established here.

\section*{Identification} \label{sec:identification}
We now return to the marginal causal risk ratio among the vaccinated, $\Psi$, and formalize our identifiability conditions.

\subsection*{Identifiability conditions} \label{sec:conditions}
    To identify the causal risk ratio among the vaccinated, $\Psi$, we rely on the following conditions
\begin{enumerate}[label=\upshape(A\arabic*), ref=A\arabic*]
    \item\label{ass1} \textit{Consistency of potential outcomes}. For all individuals $i$ and for $v \in \{0, 1\}$, we have $Y_i^v = Y_i$ when $V_i = v$. 
    \item\label{ass2} \textit{No effect of vaccination on testing negative and symptomatic among the vaccinated}. That is, $\Pr[Y^0=1 | V = 1, X] = \Pr[Y^1=1 | V = 1, X].$
    \item\label{ass3} \textit{Odds ratio equi-confounding}. Degree of unmeasured confounding bias on the odds ratio scale is the same for test-positive and test-negative illnesses, i.e. 
    $$OR_2(X) = OR_1(X), $$
    $$ \text{where } OR_y(X) = \frac{\Pr[Y^0 = y | V = 1, X]\Pr[Y^0 = 0 | V = 0, X]}{\Pr[Y^0 = 0 | V = 1, X]\Pr[Y^0 = y | V = 0, X]}.$$
    \item\label{ass4} \textit{Overlap of vaccination among test-positives and test-negatives}. Define $\mathcal{S}_y(v)$ as the support of the law of $(Y^0 = y, V = v, X)$, then for $v$ in $\{0,1\}$, then it must be that $\mathcal{S}_2(1) \subseteq \mathcal{S}_2(0)$ and $\mathcal{S}_2(v) \subseteq \mathcal{S}_1(v).$
    %\item[(A5)] No direct effect of vaccination on test-seeking behavior among the vaccinated. That is, for $i$ in $\{1,2\}$, $\Pr[T^1 = 1 | I^1 = i, V = 1, X] = \Pr[T^0 = 1 | I^0 = i, V = 1, X].$
\end{enumerate}

Assumption \ref{ass1} is a well-known identifiability condition discussed in more detail elsewhere \cite{hernan_causal_2020}. Assumption \ref{ass2} has two implications. First, it implies that the vaccine does not offer any cross-protection against other types of infection which may cause the same symptoms, including any short-term, nonspecific protection via activation of the immune system. Second, it requires that an individual's decision to seek care and get tested is unaffected by their vaccination status, though it may depend on other measured and unmeasured factors. In theory, Assumption \ref{ass2} could be evaluated in a randomized trial by comparing the incidence of testing negative (or testing positive for specific non-target pathogens) across arms. However, in practice, trials may not be adequately powered to rule out small effects. Both aspects of Assumption \ref{ass2} have been invoked in previous literature on analysis of TND studies \cite{jackson_test-negative_2013,feng_assessment_2017,schnitzer_estimands_2022}. In \ref{sec:de_testing}, we show that Assumption \ref{ass2} can be further relaxed if the effect of vaccination on testing is equivalent for test-positive and test-negative illnesses and an alternative estimand is considered.

Assumption \ref{ass3} is our key assumption and an alternative to the strict no unmeasured confounding (i.e. conditional independence) assumption suggested previously \cite{schnitzer_estimands_2022}. It states that the degree of unmeasured confounding on the odds ratio scale is the same for test-positive and test-negative illnesses. Notably, it does not require unmeasured confounders to be binary or even related to health-seeking specifically, although health-seeking behavior is likely a credible candidate for satisfying this condition. \citeauthor{lewnard_measurement_2018} mentioned a similar condition previously, although not in a formal causal framework \cite{lewnard_measurement_2018}. The assumption is also similar to a recently suggested scale-independent alternative to the parallel trend assumption in the difference-in-differences literature \cite{park_universal_2023,tchetgen_universal_2023}. We discuss the underlying parameterization in more detail in \ref{sec:or_model}. In  \ref{sec:mechanisms}, we also explore potential structural sources of equi-confounding ---including health-seeking behavior and immortal time--- as well as example violations, such as correlated vaccine-seeking  \cite{payne_impact_2023} and at home testing \cite{qasmieh2024magnitude} and how these may be resolved. When infections are mutually exclusive, Assumption \ref{ass3} simplifies to risk ratio equivalence in Equation \ref{eqn:proxy}.
Further, recalling $Y = I \times T$, we can split Assumption \ref{ass3}  into:
\begin{enumerate}[label=\upshape(A3\alph*), ref=A3\alph*]
    \item\label{ass3a}  \textit{Odds ratio equi-confounding}. Degree of unmeasured confounding bias on the odds ratio scale is the same for symptomatic illness regardless if $I=1$ or $I=2$ is cause, i.e. 
    $$\frac{\Pr[I^0 = 2 | V = 1, X]}{\Pr[I^0 = 2 | V = 0, X]} =\frac{\Pr[I^0 = 1 | V = 1, X]}{\Pr[I^0 = 1 | V = 0, X]}.$$
    \item\label{ass3b} \textit{Odds ratio equi-selection}. Degree of unmeasured selection bias, e.g. receiving a test were the subjects not vaccinated, on the odds ratio scale is the same for all symptomatic illness regardless if $I=1$ or $I=2$ is cause, i.e. 
    $$\frac{\Pr[T^0 = 1 | I^0 = 2, V = 1, X]}{\Pr[T^0 = 1 | I^0 = 2, V = 0, X]} =\frac{\Pr[T^0 = 1 | I^0 = 1, V = 1, X]}{\Pr[T^0 = 1 | I^0 = 1, V = 0, X]}.$$
\end{enumerate}
Assumptions \ref{ass3a} and \ref{ass3b} are technically stronger than \ref{ass3} alone as \ref{ass3} allows, in principle, for situations in which the terms in Assumptions \ref{ass3a} and \ref{ass3b} are both unequal but  cancel when multiplied. 

Finally, Assumption \ref{ass4} is similar to, although less restrictive than, the well-known positivity condition in causal inference \cite{hernan_causal_2020}. Specifically, Assumption \ref{ass4} requires overlap of vaccination propensity between the observed vaccinated and unvaccinated among the test positive cases as well as overlap of vaccination propensity across test-positive cases and test-negative controls within observed vaccination groups. 

The causal directed acyclic graphs (DAG) in Figure \ref{fig:dag} shows a possible mechanism satisfying the identifiability conditions above.

\subsection*{Identification in a cohort study}
Our first proposition shows that, under  Assumptions \ref{ass1} - \ref{ass4}, $\Psi$ is identifiable under cohort sampling. 
\begin{proposition}\label{prop1}
    Given Assumptions \ref{ass1} - \ref{ass4}, $\Psi$ is identifiable under cohort sampling, i.e. $O_{cohort} = \{(X_i, V_i, Y_i)\}_{i=1}^N,$ by the expressions 
    \begin{equation}\label{eqn:om_estimand}
        \Psi_{om} \equiv \dfrac{\E[V \mathbbm 1 (Y = 2)]}{\E\left[V \mathbbm 1(Y = 1) \dfrac{\Pr[Y = 2 | V = 0, X]}{\Pr[Y = 1 | V = 0, X]}  \right]}
    \end{equation}
    and 
    \begin{equation}\label{eqn:ipw_estimand}
        \Psi_{ipw} \equiv \dfrac{\E[V \mathbbm 1 (Y = 2)]}{\E\left[ (1 - V) \mathbbm 1(Y = 2) \dfrac{\Pr[V=1|Y = 1, X]}{\Pr[V=0|Y = 1, X]}\right]}.
    \end{equation}
    % \begin{equation}\label{eqn:ipw_estimand}
    %     \Psi_{ipw} \equiv \dfrac{\E[V \mathbbm 1 (Y = 2)]}{\E\left[ (1 - V) \mathbbm 1(Y = 2) \dfrac{\pi_1(X)}{1 - \pi_1(X)}\right]}
    % \end{equation}
    %where $\pi_y(X) = \Pr[V = 1| Y = y, X]$ and $\mu_v(X) = \Pr[Y = 2$.
    \end{proposition}
    \begin{proof}
        Proof of Proposition 1 is deferred to \ref{sec:proofs}.
    \end{proof}

Expression \ref{eqn:om_estimand} is a generalization of the result for $\Psi_{om}(X)$ discussed earlier, where the de-biasing term is now standardized over the distribution of $X$ among the vaccinated. Expression \ref{eqn:ipw_estimand} offers an alternative representation of the de-biasing term based on Assumptions \ref{ass1} - \ref{ass4} and the invariance property of the odds ratio, as 
$$\frac{\Pr[Y^0 = 2| V = 1, X]}{\Pr[Y^0 = 2 | V = 0, X]} = \frac{\Pr[V = 1 | Y^0 = 2,  X]}{\Pr[V = 0 | Y^0 = 2, X]} =\frac{\Pr[V = 1 | Y = 1, X]}{\Pr[V = 0 | Y = 1, X]}.$$
Following standard terminology \cite{robins_estimating_1992}, we refer to Expression \ref{eqn:om_estimand} as the \textit{outcome-modeling} estimand and Expression \ref{eqn:ipw_estimand} as the \textit{inverse-probability weighting} estimand.

\begin{remark}
The quantity $\Pr[V = 1 | Y^0 = 2, X]$ is sometimes referred to as the \textit{extended propensity score} function, as it extends the standard propensity score to allow for unmeasured confounding by conditioning on the treatment-free potential outcome --- in this case $Y^0 = 2$. As shown previously \cite{tchetgen_single_2023}, in the presence of unmeasured confounding, knowledge of the extended propensity score is sufficient to identify treatment effects among the treated. Similarly knowledge of $\Pr[V = 1 | Y^1 = 2, X]$ would permit identification of treatment effects among the untreated, and knowledge of both would permit the identification of the overall effect \cite{tchetgen_single_2023}.
\end{remark}

\begin{remark}
Recent work by \textcite{li2024comparison} compares the performance of the TND with analyses in the full cohort using target trial emulation. However, it is unclear whether these designs target the same estimand. Indeed, Expressions \ref{eqn:or_estimand}, \ref{eqn:om_estimand}, and \ref{eqn:ipw_estimand} suggest an alternative analysis in the cohort ---essentially a difference-in-differences analysis comparing test-positive and test-negative outcomes--- would target $\Psi$ and could offer a more direct comparison with the TND estimators introduced in this paper. We also note that, in a cohort study, alternative estimates of absolute risk differences could be derived under equi-confounding.
\end{remark}

\subsection*{Identification under outcome-dependent sampling in a TND}

Expressions \ref{eqn:om_estimand} and \ref{eqn:ipw_estimand} require absolute probabilities that cannot be estimated under the outcome-dependent sampling scheme of the TND. Nonetheless, the next proposition shows that $\Psi$ remains identifiable 
 \begin{proposition}\label{prop2}
      Under test-negative sampling, i.e. $O_{TND} = \{(X_i, V_i, S_i=1, Y_i)\}_{i=1}^n,$ with selection $S = \mathbbm 1(Y\neq 0)$, $\Psi_{om}$ and $\Psi_{ipw}$ are equivalent to 
    \begin{equation}\label{eqn:om_estimand_tnd}
        \Psi_{om}^* = \dfrac{\E[V \mathbbm 1 (Y = 2)|S =1]}{\E\left[V\mathbbm 1 (Y = 1) \dfrac{\Pr[Y = 2 | S = 1, V = 0, X]}{\Pr[Y = 1| S = 1, V = 0, X]}\Big| S = 1 \right]}
    \end{equation}
    and 
     \begin{equation}\label{eqn:ipw_estimand_tnd}
        \Psi_{ipw}^* = \dfrac{\E[V \mathbbm 1 (Y = 2)|S =1]}{\E\left[ (1 - V) \mathbbm 1 (Y = 2) \dfrac{\Pr[V=1 | S=1, Y=1, X]}{\Pr[V=0 | S=1, Y=1, X]} \bigg| S = 1\right]}.
    \end{equation}
    Therefore by Proposition \ref{prop1} and assuming conditions \ref{ass1}-\ref{ass4} hold, $\Psi_{om}^*$ and $\Psi_{ipw}^*$ are also identifying expressions for $\Psi$.
 \end{proposition}
 \begin{proof}
    Proof of Proposition 2 is deferred to \ref{sec:proofs}.
 \end{proof}
    
 Expressions \ref{eqn:om_estimand_tnd} and \ref{eqn:ipw_estimand_tnd} only require conditional probabilities among the tested and thus are identifiable in a TND. As before, we refer to Expression \ref{eqn:om_estimand_tnd} as the outcome-modeling estimand of the TND and Expression \ref{eqn:ipw_estimand_tnd} as the inverse-probability weighting estimand of the TND.

 \begin{remark}
     Comparing Expressions \ref{eqn:om_estimand} and \ref{eqn:om_estimand_tnd}, we note that the ``sampling scheme'' of the TND can be viewed as outcome-dependent or case-control sampling of an outcome of interest ($Y=2$) and a corresponding negative outcome control ($Y=1$). 
 \end{remark}

 \begin{remark}
     Expressions \ref{eqn:om_estimand_tnd} and  \ref{eqn:ipw_estimand_tnd} resemble those proposed assuming unconfoundedness \cite{jiang_tnddr_2023}. In that work, the authors introduce a class of marginal estimands in a TND that can be written as 
     \begin{equation*}
         \dfrac{\E\left[\Pr[Y=2 | V = 1, X = x, S = 1]\omega_1(x) | S = 1 \right]}{\E\left[\Pr[Y=2 | V = 0, X = x, S = 1]\omega_0(x) | S = 1\right]}
     \end{equation*}
     where $\omega_v(x)$ are ``de-biasing weights'' used to adjust for bias due to outcome-dependent sampling. They describe weights based on both outcome-modeling and inverse probability weighting. We note that the denominators of Expressions \ref{eqn:om_estimand_tnd} and  \ref{eqn:ipw_estimand_tnd} both have a similar form, using alternative weights 
     \begin{equation*}
         \omega_0(x) \equiv \dfrac{\Pr[Y = 1 | S = 1, V = 1, X]}{\Pr[Y=1| S = 1, V = 0, X]} = \dfrac{\Pr[V=1 | S = 1, Y = 1, X]}{\Pr[V=0| S = 1, Y = 1, X]}
     \end{equation*}
     However, these weights simultaneously address outcome-dependent sampling and adjust for unmeasured equi-confounding. 
 \end{remark}

\section*{Estimation}\label{sec:estimation}
We now describe estimation strategies for the causal risk ratio among the vaccinated in a TND, based on the identifying expressions in Proposition \ref{prop2}. However, we describe in \ref{sec:app_estimation} estimators based on the results in Proposition \ref{prop1} in a cohort study. For convenience, we define an indicator for test positive case in the TND sample $Y^* = \mathbbm 1 (Y = 2)$.

Traditionally in a TND, the analyst performs a logistic regression of $Y$ conditional on $X$ and $V$, such as:
$$\log \left\{\dfrac{\Pr[Y^*=1 | S = 1, V, X]}{\Pr[Y^* = 0 | S = 1, V, X]}\right\} = \gamma_0 + \gamma_1 V + \gamma_2^\prime X,$$
where $\gamma_1$ is the focal coefficient. As we have shown, $\exp(\gamma_1)$ is equal to $\Psi^*_{om}(X)$, and by extension, the conditional causal risk ratio among the vaccinated, $\Psi(X)$, under identifiability Assumptions \ref{ass1} - \ref{ass4} and correct model specification. 

Alternatively, Expressions \ref{eqn:om_estimand_tnd} and \ref{eqn:ipw_estimand_tnd} suggest two plug-in estimators valid under effect modification. Expression \ref{eqn:om_estimand_tnd} suggests the plug-in estimator:
\begin{equation}\label{eqn:om_estimator}
    \widehat{\Psi}_{om}^* = \dfrac{\sum_{i=1}^n V_i Y^*_i}{\sum_{i=1}^n V_i (1 - Y_i^*)\dfrac{\widehat{\mu}^*_0(X_i)}{1 - \widehat{\mu}^*_0(X_i)}},
\end{equation}
where $\mu^*_v(X) = \Pr[Y^* =1 \mid S=1, V=v, X]$ is the probability of testing positive among those with vaccine status $V =v$ in the TND sample. This could be obtained, for instance, via a logistic regression of $Y^*$ conditional on $X$ and $V$.

Expression \ref{eqn:ipw_estimand_tnd}, on the other hand, suggests the plug-in estimator:
\begin{equation}\label{eqn:ipw_estimator}
    \widehat{\Psi}_{ipw}^* = \dfrac{\sum_{i=1}^n V_i Y^*_i}{\sum_{i=1}^n (1 - V_i) Y^*_i \dfrac{\widehat\pi^*_0(X_i)}{1 - \widehat\pi^*_0(X_i)}},
\end{equation}
where $\pi_y(X) = \Pr[V=1\mid S=1, Y^*=y, X]$ is the probability of vaccination among the test-negative controls. This probability could be obtained, for instance, via a logistic regression of $V$ on $X$ among those with $Y^*=0$. In \ref{sec:app_estimation}, we discuss methods for estimating standard errors and confidence intervals for $\widehat{\Psi}_{om}^*$ and $\widehat{\Psi}_{ipw}^*$ via bootstrapping or stacked estimating equations. 

The first estimator, $\widehat{\Psi}_{om}^*$, requires  correct specification of the outcome model, while the second estimator, $\widehat{\Psi}_{ipw}^*$, requires correct specification of the extended propensity score model. %The preferred estimator may depend on the context. For instance, if more is known about the assignment mechanism for vaccination, the weighting estimator may be preferred \cite{robins_estimating_1992,braitman_rare_2002}. Conversely, if the process leading to infection and testing is better understood, the outcome estimator may be more appropriate. 
In practice, however, both models may be difficult to specify correctly. To address this, we construct a doubly robust estimator for $\Psi$ that is consistent and asymptotically normal if the model for either $\mu^*_0(X)$ or $\pi^*_0(X)$ or both are correctly specified. Full details of the doubly robust approach are included in \ref{sec:eif}, but are based on noting that an equivalent expression of $\Psi_{om}$ and $\Psi_{ipw}$ is
\begin{equation*}
    \dfrac{\E[V \mathbbm 1 (Y = 2)|S =1]}{\E\left[V\mathbbm 1 (Y = 2) \exp\{-\phi^*(X)\}\Big| S = 1 \right]},
\end{equation*}
where 
\[\phi^*(X) = \log \dfrac{\Pr[Y=2|S=1, V=1,X]\Pr[Y=1|S=1, V=1,X]}{\Pr[Y=1|S=1, V=0,X]\Pr[Y=2|S=1, V=0,X]}\]
is the log of the conditional odds ratio function. As described in \textcite{tchetgen_tchetgen_doubly_2010}, a doubly robust estimator of $\alpha^*(X)$ may be obtained as the solution to the empirical analogue of the population moment equation
\[E[(1,X)'(V-\pi_0^*(X))\exp\{-\phi^*(X)VY^*\}(Y^* - \mu_0^*(X))] = 0.\]
Let $\widehat{\phi}_{dr}^*(X)$ be one such solution, a doubly robust estimator of $\Psi$ is then
\begin{equation}\label{eqn:dr_estimator}
    \widehat{\Psi}_{dr}^* = \dfrac{\sum_{i=1}^n V_i Y^*_i}{\sum_{i=1}^nV_iY^*_i \exp\{\widehat{\phi}_{dr}^*(X)\}}.
\end{equation}

\section*{Sensitivity analysis}
In practice, the odds ratio equi-confounding assumption may not hold such that
$$OR_2(X) \neq OR_1(X).$$ 
However, the decomposition in Expression \ref{eqn:decomposition} suggests a sensitivity analysis that allows for departures from equi-confounding as follows. Consider the scenario where the odds ratio function $OR_2(X)$ is instead equal to:
\begin{equation}\label{eq:sensitivity-analysis}OR_2(X) = e^{\eta q(X)} OR_1(X), \end{equation}
where $\eta$ is a scalar and $q(X)$ is a user-specified  monotone sensitivity function. Together, they parameterize the deviation from equi-confounding on the odds ratio scale. This is an example of the well-known exponential tilt model \cite{scharfstein_adjusting_1999, liu_identification_2020}. Setting $\eta = 0$ corresponds to the case where equi-confounding holds, with values of $\eta$ further from zero representing larger violations. For example, one might choose $q(X) = 1$, suggesting that confounding occurs uniformly across levels of $X$, and then vary $\eta$ over a grid of values in the interval $(-\omega, \omega)$.

% With similar derivation to Section~\ref{sec:effect_among_vaccinated}, it can be shown that under Equation~\eqref{eq:sensitivity-analysis}, the conditional causal risk ratio $\Psi(X)$ can be identified as
% \begin{equation}
%     \dfrac{\Pr[I = 2, T = 1 | V = 1, X]/\Pr[I = 1, T = 1 | V = 1, X]}{\Pr[I = 2, T = 1 | V = 0, X]/\Pr[I = 1, T = 1 | V = 0, X]}e^{-\eta q(X)}.
% \end{equation}
% That is, the conditional causal risk ratio among the vaccinated and the odds ratio estimated in the test-negative study sample differ by a factor of $e^{-\eta q(X)}$. Furthermore, the marginal risk ratio among the vaccinated $\Psi$ can be identified as 

% $$\dfrac{\Pr(I^* = 1 | S = 1, V = 1)}{E\left\{  \dfrac{\Pr(I^* = 0 | S = 1, V = 1, X)}{\Pr(I^* = 0| S = 1, V = 0, X)}\Pr(I^* = 1 | S = 1, V = 0, X)e^{\eta q(X)} \Big| S = 1, V = 1 \right\}}.$$
% If $q(x)=q_0$ is a constant, then $\Psi=\Psi_{om}^*e^{-\eta q_0}$, that is, the estimated marginal causal risk ratio among vaccinated under Assumptions A1-A5 will also differ by a factor of $e^{\eta q_0}$.

\section*{Simulation}

To illustrate a data generation process where our assumptions hold and to evaluate the empirical performance of the proposed estimators when they do not, we conduct a simulation study. The data generation process is as follows:
\begin{align*}
    X, U &\sim \text{Unif}(0,1)\\
    V\mid X, U & \sim \text{Bernoulli}(\expit(\alpha_0 + \alpha_X X + \alpha_U U))\\
    I^v \mid V, X, U &\sim \text{Multinomial}(1-p_1(v, X, U) - p_2(v, X, U), p_1(v, X, U), p_2(v, X, U))\\
    T^v\mid I^v=i, V, X, U &\sim \text{Bernoulli}(\mathbbm 1(i>0)\exp\{(\tau_{1} + \tau_{1V}v) \mathbbm 1(i=1) + (\tau_{2} + \tau_{2V} v + \tau_{2U} U ) \mathbbm 1(i=2) \\&\qquad \qquad\qquad + \tau_X X + \tau_U U \})
\end{align*}
where $U$ is an unmeasured confounder and
\begin{align*}
    p_1(v, X, U) & = \Pr[I^v = 1 | V, X, U] = \exp(\beta_{10} + \beta_{1V}v + \beta_{1X}X + \beta_{1VX}vX + \beta_{1U}U) \\
    p_2(v, X, U) & = \Pr[I^v = 2 | V, X, U] = \exp(\beta_{20} + \beta_{2V}v + \beta_{2X}X + \beta_{2VX}vX + \beta_{2U}U).
\end{align*} We apply $Y = I \times T$. Under this process, potential outcomes $Y^v$ for $v=0,1$ and $y=1, 2$ are generated from:
\begin{align*}
    \Pr[Y^v = y\mid V, X, U] &= p_y(v, X, U)\exp\{\tau_y + \tau_{yV} v + \tau_X X + \tau_U U + \tau_{2U}U\mathbbm 1(y=2) \}.
\end{align*}
This implies the conditional independence $Y^v \indep V\mid X, U$ holds for $v=0,1$, meaning that conditioning on $U$ and $X$ is sufficient to control confounding. When there is no effect modification by $U$ or $X$ ($\beta_{2VX}=0$) and no direct effect on testing for focal illness ($\tau_{2V} = 0$), the causal risk ratio is
\begin{align*}
    \Psi &= \exp(\beta_{2V}).
\end{align*}
Finally, under TND sampling with $S=\mathbbm{1}(Y \neq 0)$, it can also be shown that
\begin{align*}
    \Pr[Y=2 \mid  S=1, V, X, U] &= \expit\{(\beta_{20}-\beta_{10}+\tau_2 - \tau_1) + (\beta_{2V} - \beta_{1V} + \tau_{2V} - \tau_{1V}) V\\ &\qquad\qquad+(\beta_{2X}-\beta_{1X})X+ (\beta_{2VX} - \beta_{1VX})VX + (\beta_{2U} - \beta_{1U} + \tau_{2U})U\},
\end{align*}
which follows a logistic regression model.

Violations of Assumption \ref{ass3} are controlled via $\beta_{2U}$, $\beta_{1U}$ and $\tau_{2U}$. Under our set up, we have:
\begin{align*}
    \dfrac{\Pr[Y^0=2 \mid V=1,X]}{\Pr[Y^0=2\mid V=0,X]}=\dfrac{\E[\exp\{(\tau_U  + \tau_{2U}+\beta_{2U})U\}\mid V=1, X]}{\E[\exp\{(\tau_U  + \tau_{2U}+\beta_{2U})U\}\mid V=0, X]},
\end{align*}
and 
\begin{align*}
    \dfrac{\Pr[Y^0=1 \mid V=1,X]}{\Pr[Y^0=1\mid V=0,X]}=\dfrac{\E[\exp\{(\tau_U  +\beta_{1U})U\}\mid V=1, X]}{\E[\exp\{(\tau_U +\beta_{1U})U\}\mid V=0, X]},
\end{align*}
implying that that Assumption \ref{ass3} holds when $\beta_{2U}=\beta_{1U}$ and $\tau_{2U}=0$ (Note that under these conditions the logistic model above similarly does not depend on $U$). Following the previous discussion, Assumption \ref{ass3} can be viewed as composed of two sub conditions (\ref{ass3a} and \ref{ass3b}): First, the unmeasured confounder exerts an equivalent effect on $I=1$ and $I=2$ (i.e. $\beta_{1U}=\beta_{2U}$), and second, there is no interaction between the unmeasured confounder and the source of illness for the probability of testing on the multiplicative scale (i.e. $\tau_{2U}=0$). 

Likewise, violations of Assumption \ref{ass2} are controlled by $\beta_{1V}$ and $\tau_{1V}$. When $\beta_{1V} \neq 0$ there is a direct effect of vaccination on test-negative illness $I = 1$ and when $\tau_{1V} \neq 0$ there is a direct effect of vaccination on testing among those with test-negative illness. When this occurs, we can at most identify the ratio of vaccine effects on the two illness outcomes, i.e. $\exp(\beta_{2V} + \tau_{2V})/\exp(\beta_{1V} + \tau_{1V})$, provided there is no effect modification by $X$ or $U$. As we show in \ref{sec:de_testing} and illustrate further via simulation below, under the special case of equal effects on testing, i.e. $\tau_{2V} = \tau_{1V}$, certain effects on symptomatic illness may still be recovered.  
%Finally, under this setup, the causal risk ratio for \subsection*attended illness is 
% \begin{align*}
%     \Psi &= \exp(\beta_{2V} + \tau_{2V})\dfrac{E[\exp\{(\beta_{2X}+\beta_{2VX})X\}\mid V=1]}{E[\exp(\beta_{2X}X)\mid V=1]}.
% \end{align*}

 We generate a target population of $N = 15,000$ resulting in TND samples of test-positive cases and test-negative controls of between $1000$ and $2000$. We consider eight scenarios: 
 \begin{enumerate}
    \item No unmeasured confounding
    \item Unmeasured confounding, but Assumptions \ref{ass1} - \ref{ass4} hold
    \item Direct effect of vaccination on $I=1$
    \item Equi-confounding is violated
    \item Equi-selection is violated
    \item Equal effects of vaccination on testing
    \item Unequal effects of vaccination on testing
    %\item Infections $I=1$ and $I=2$ are not mutually exclusive.
    \item All assumptions hold but there is effect modification by covariates $X$
 \end{enumerate}
 Parameter values for all scenarios are shown in eTable \ref{tab:simparams} and descriptions of the estimation methods are provided in eTable \ref{tab:methods}. For each scenario, we generate 1000 replicates and estimate $\Psi$ using the proposed estimators $\widehat{\Psi}_{om}^*$, $\widehat{\Psi}_{ipw}^*$, and $\widehat{\Psi}_{dr}^*$ as well as the conventional logistic regression estimator, $\widehat{\Psi}_{om}^*(X)$, and calculate the bias and coverage of 95\% confidence intervals based on the true value. For comparison, we also consider the following estimators based on data from the full cohort: a conventional outcome-modeling estimator adjusting for $X$ alone, another conditioning on $X$ and $U$ (covering the hypothetical case that all confounders are measured), and one based on $\Psi_{om}$ in Proposition \ref{prop1}. Finally, in scenario 8, we demonstrate the double robustness property of $\widehat{\Psi}_{dr}^*$ by adjusting the data generation process for $V$ and $I^v$ to include squared terms in $X$. Additional details on the setup and precise estimation methods are provided in  \ref{sec:moresim}.

\begin{figure}
    \centering
    \includegraphics{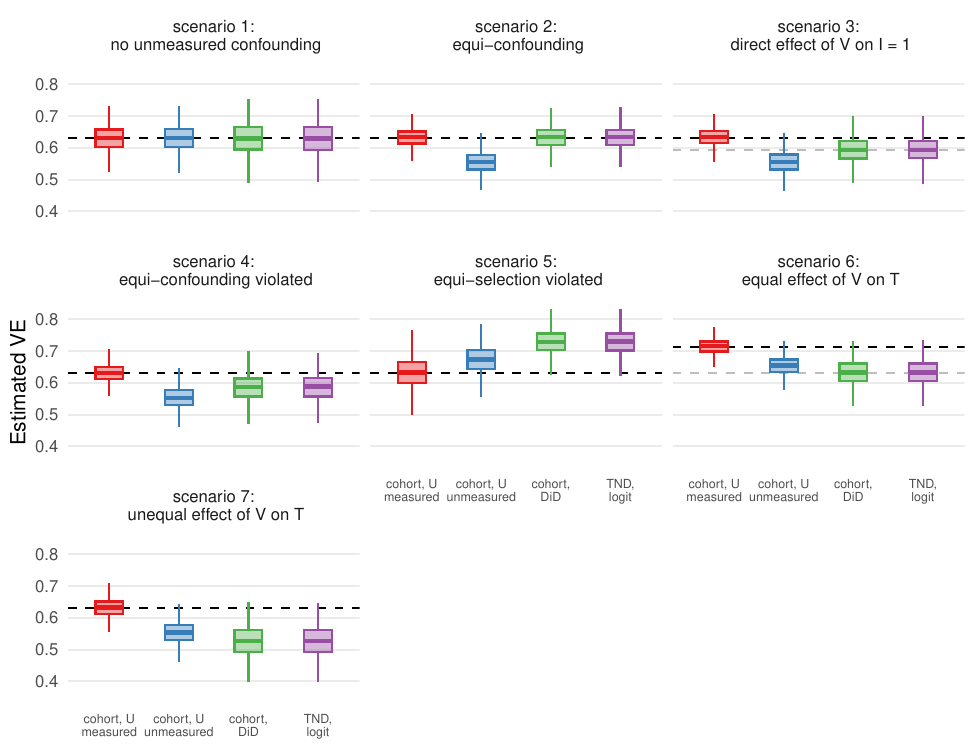}
    \caption{
        Bias of vaccine effectiveness (VE) estimators across different simulation scenarios. In each scenario, we compare: (i) ideal cohort estimator with $U$ measured, (ii) more typical cohort estimator with $U$ unmeasured, (iii) a difference-in-differences (DiD) estimator in the cohort based on equation \ref{eqn:om_estimand}, and (iv) the conventional TND estimator based on logistic regression $\widehat{\Psi}_{om}^*(X)$. Results are based on 1000 Monte Carlo replications. Black dashed line is the true VE. In scenario 3, grey dashed line is the ratio of VEs for test-positive versus test-negative illness. In scenarios 6 and 7, grey dashed line VE against symptomatic illness (for more on this see  \ref{sec:de_testing}).}
    \label{fig:sims1}
\end{figure}

\begin{figure}
    \centering
    \includegraphics{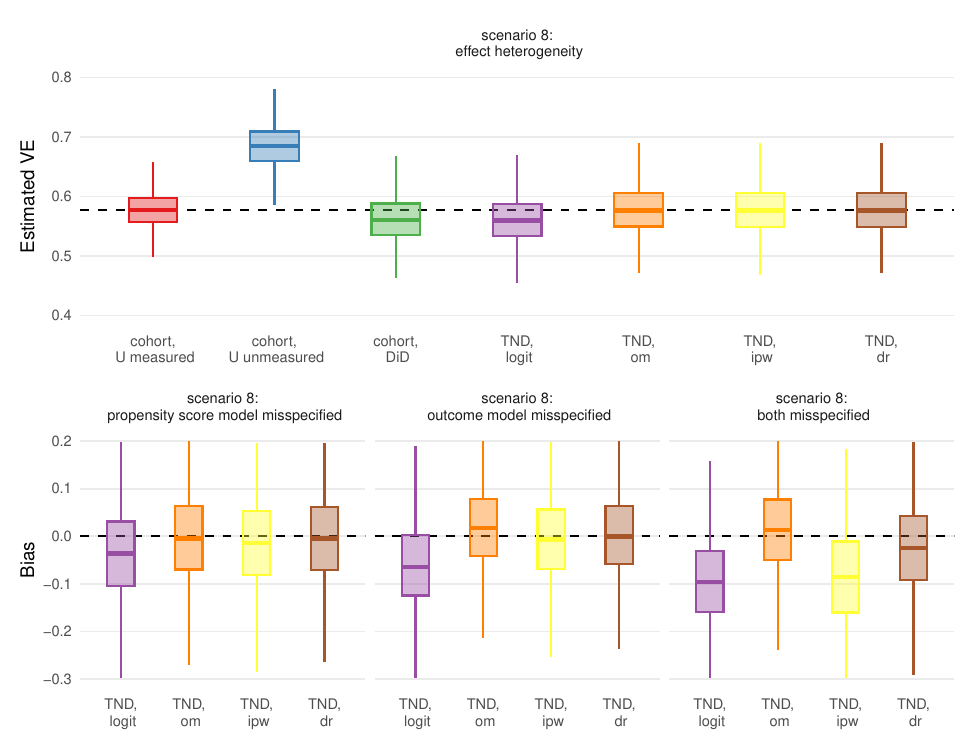}
    \caption{Bias of vaccine effectiveness (VE) estimators under effect heterogeneity and misspecification of the nuisance functions. We add to the estimators considered in Figure \ref{fig:sims1} the novel estimators of the marginal RRV introduced in section \ref{sec:estimation}: the outcome modeling estimator (TND, om), the inverse probability weighting estimator (TND, ipw), and the doubly robust estimator (TND, dr). Results are based on 2000 Monte Carlo replications. Black dashed line is the true VE.}
    \label{fig:sims2}
\end{figure}

The results are reported in Figures \ref{fig:sims1} and \ref{fig:sims2} with further details in eTables \ref{tab:sims} and \ref{tab:sims2}. When there is no effect modification (scenarios 1 to 7) the proposed TND estimators and traditional logistic regression estimator are the same, so we show only the results from the latter, $\widehat{\Psi}_{om}^*(X)$, under the name ``TND, logit''. When there is no unmeasured confounding, all estimators are unbiased (scenario 1). In all other scenarios, the cohort estimator without $U$ is biased. When there is unmeasured confounding but our assumptions hold, the TND estimator is unbiased (scenario 2). When there is an effect of vaccination on the test-negative illness (scenario 3), the TND estimator is biased towards the null (when the effect is also protective); however, it does identify the ratio of vaccine effects against $I = 2$ versus $I = 1$ (grey dashed line). When equi-confounding or equi-selection are violated the TND estimator is biased (scenarios 4 and 5).  When there is a direct effect of vaccination on testing behavior, the TND estimator is unbiased for the risk ratio against symptomatic illness (grey dashed line) if the effect is equivalent for test-negative and test-positive illnesses (scenario 6, for more on this see  \ref{sec:de_testing}), but is biased when they are unequal (scenario 7). In all scenarios, the estimates from the TND estimator are the same as the difference-in-differences estimator (cohort, DiD) based on Proposition \ref{prop1} applied to the full cohort even though the TND estimator uses only information among the tested, demonstrating how TND may be viewed as essentially a more efficient sampling design. eTable \ref{tab:sims} shows that, when unbiased, confidence intervals derived from estimating equations in  \ref{sec:app_estimation} exhibit nominal coverage.

Finally, in the more realistic scenario, that the effect of vaccination is heterogeneous, the proposed TND estimators, $\widehat{\Psi}_{om}^*$,$\widehat{\Psi}_{ipw}^*$, and $\widehat{\Psi}_{dr}^*$, remain unbiased for the marginal estimand while the conventional logistic regression approach, $\widehat{\Psi}_{om}^*(X)$, is biased (scenario 8). In Figure \ref{fig:sims2}, we confirm the double robustness property of $\widehat{\Psi}_{dr}^*$. As expected, when the outcome model is misspecified $\widehat{\Psi}_{ipw}^*$ and $\widehat{\Psi}_{dr}^*$ remain unbiased while $\widehat{\Psi}_{om}^*$ is biased. Likewise, when the extended propensity score model is misspecified $\widehat{\Psi}_{om}^*$ and $\widehat{\Psi}_{dr}^*$ remain unbiased while $\widehat{\Psi}_{ipw}^*$ is biased. When both models are misspecified all estimators are biased.

\section*{Discussion} \label{sec:discussion}

We have shown that the TND can yield unbiased estimates of vaccine effectiveness among the vaccinated in the presence of unmeasured confounding provided certain conditions are met: namely, the vaccine does not affect the test-negative illness and the net effect of unmeasured confounders  on symptomatic illness and testing is equal for both test-positive and test-negative infections. While our results apply to the commonly used logistic regression estimator, we also derived novel estimators based on outcome modeling and inverse probability weighting as well as a doubly robust estimator that is consistent if one or both of the outcome model and propensity score model are correctly specified.

The plausibility of the equi-confounding assumption may be increased by knowledge of the source of the test-negative illness.  In previous TNDs for influenza that included additional testing to confirm the source of test-negative illness the most common causes were rhinovirus, respiratory syncytial virus, human metapneumovirus, parainfluenza virus, bocavirus, coronavirus, adenovirus, enterovirus and others \cite{chua_use_2020-1}. Ideally, the test-negative illness should have a similar exposure mechanism and clinical presentation as the target illness but remain unaffected by the vaccine. For example, in pneumococcal disease, non-vaccine serotypes of \textit{Streptococcus pneumoniae} serve as ideal test-negative controls \cite{broome_pneumococcal_1980}. However, more careful selection of test-negative infections based on source pathogen is currently limited by infrequent use of concurrent or multiplex testing. Care must also be taken when test-negative infections have associated vaccines as these can be sources of \textit{unequal} confounding if not adjusted for.

Even when equi-confounding does not hold, as our simulations suggest, the TND may still reduce bias relative to other approaches based only on a minimal covariate set, when test-negative illness serves as a reasonable, if imperfect, proxy for residual confounding. We also propose a sensitivity analysis that allows for straightforward departures from equi-confounding via the exponential tilt model. Alternatively, additional negative controls can be leveraged in a proximal inference framework to identify effects under more general unmeasured confounding structures \cite{li_double_2023}.

Our results also depend on the assumption that the vaccine does not affect other infections or influence test-seeking behavior (except through infection). The former may be violated if vaccination provides short-term nonspecific protection through immune activation. Previous research has suggested that this is possible for influenza vaccines \cite{cowling_increased_2012}. However, other results have shown the opposite \cite{sundaram_influenza_2013}. Violations of this assumption would bias TND estimates, though in some cases, if the effect is in the same direction, the TND might still provide a lower bound on true vaccine effectiveness. If vaccination exerts equal effects on test-seeking behavior for test-positive and test-negative infections, as our simulations demonstrate, estimates of vaccine effectiveness against symptomatic infection remain unbiased. 

%While our focus has been on outpatient TNDs, where voluntary care-seeking is a major factor, TNDs in inpatient settings—particularly during the COVID-19 pandemic—warrant careful consideration. The composition of test-negative illnesses and the nature of unmeasured confounding may differ in these settings.

Last, our results have implications for other study designs for vaccine evaluation. First, \textcite{wang_randomization_2022} recently proposed a cluster-randomized TND that corrects for potential bias caused by intervention-induced differential health-seeking behavior in an unblinded trial. In our framework, this corresponds to a setting where Assumption \ref{ass3a} holds by design, as vaccination is randomized, but Assumption \ref{ass3b} does not, as individuals are aware of their assignment and therefore may differentially seek care when sick. As mentioned above, to remove the influence of intervention effects on health-seeking requires them be equal across test-positive and test-negative illnesses on the multiplicative scale. Second, in settings where full cohort data are available, comparing test-positive and test-negative infections via Expressions \ref{eqn:om_estimand} or \ref{eqn:ipw_estimand} offers an alternative approach to confounding control, allowing for estimation of absolute risks under different vaccination regimes.

\newpage

%TC:ignore

\printbibliography

    \begin{figure}[p]
        \centering
        \begin{tikzpicture}[> = stealth, shorten > = 1pt, auto, node distance = 3cm, inner sep = 0pt,minimum size = 0.5pt,  very thick]
            \tikzstyle{every state}=[
              draw = none,
              fill = none
            ]
            \node[state] (x) {$X$};
            \node[state] (v) [right of=x] {$V$};
            \node[state] (i) [right of=v] {$I = 2$};
            \node[state] (t) [right of=i] {$T$};
            \node[state] (i1) [below of=i] {$I = 1$};
            \node[state] (u) [below of=v] {$U$};
    
            \path[->] (x) edge node {} (v);
            \path[->] (x) edge [out=45, in=135] node {} (i);
    
            \path[->] (v) edge node {} (i);
            
            \path[->] (i) edge [forestgreen, pos=0.5, sloped, above] node [font=\tiny] {\circled{A3b}} (t);
            \path[->] (i1) edge [forestgreen, pos=0.5, sloped, below] node [font=\tiny] {\circled{A3b}} (t);
    
            \path[->] (x) edge [out=45, in=135] node {} (t);

            \path[->] (u) edge node {} (x);
            \path[->] (u) edge node {} (v);
            \path[->] (u) edge [bblue, pos=0.5, sloped, above] node [font=\tiny] {\circled{A3a}} (i);
            \path[->] (u) edge node {} (t);
            \path[->] (u) edge [bblue, pos=0.5, sloped, below] node  [font=\tiny] {\circled{A3a}} (i1);
            \end{tikzpicture}
        \caption{Causal directed acyclic graph (DAG) showing the assumed relationship between the variables in a test-negative study under equi-confounding. Here we split the $I$ node into $I=1$ and $I=2$ to show how the former acts as a negative outcome control. The equi-confounding assumption itself, as a parametric restriction, is not easy to represent on a DAG using standard notation. Here, we add labels over the arrows that must represent equal effects on the odds ratio scale for Assumptions \ref{ass3a} and \ref{ass3b} to hold. }
        \label{fig:dag}
    \end{figure}
    \clearpage

\clearpage

\title{\textit{eAppendix:} Identification and estimation of vaccine effectiveness in the test-negative design under equi-confounding}
\date{ }
\maketitle
\pagenumbering{arabic}% resets `page` counter to 1

\textbf{This file contains:}

\textbf{eAppendices 1-9, eTables 1-4, eFigures 1-2}
\newpage 
\begin{appendix}
    \renewcommand{\thesection}{eAppendix \arabic{section}}
    \renewcommand{\thesubsection}{\arabic{section}.\arabic{subsection}}
    
    \renewcommand{\figurename}{eFigure}
    \setcounter{figure}{0}

    \renewcommand{\tablename}{eTable}
    \setcounter{table}{0}
    
    \renewcommand{\theequation}{A\arabic{equation}}
    \setcounter{equation}{0}

    \singlespacing
     \part{ } % Start the appendix part
    \parttoc % Insert the appendix TOC

%    \appendixwithtoc
    \newpage
    \begin{refsection}  

    \section{A nonparametric odds ratio model}\label{sec:or_model}
Following \textcite{tchetgen_universal_2023}, we parameterize a unit's contribution to the likelihood for the potential outcome under no treatment $Y^0$ conditional on $V$ and covariates $X$ and assuming independent sampling. Let
\begin{align*}
    h(y, x) &= f(Y^0 = y | V = 0, X = x), \\
    \beta(y,  x) &= \log \dfrac{f(Y^0 = y | V = 1, X = x)f(Y^0 = y_{ref} | V = 0, X = x)}{f(Y^0 = y_{ref} | V = 1, X = x)f(Y^0 = y | V = 0, X = x)},
\end{align*}
where $h(y, x) $ is the conditional density among the unvaccinated and the function $ \beta(y,  x)$ is the log of the generalized odds ratio function with $\beta(y_{ref},  x) = 0$ for user-specified reference values $y_{ref}$. Here we use $y_{ref} = 0$. Thus, $\beta(y,  x) = 0$ implies no unmeasured confounding for symptomatic infection or testing given $X = x$ and $\beta(y,  x) \neq 0$ encodes the degree of unmeasured confounding bias at the distributional level. 

This parameterization implies:
\begin{align*}
    f(Y^0 = y | V = 1, X = x) \propto h(y, x) \exp\{\beta(y, x)\}
\end{align*}
with 
\begin{align*}
    f(Y^0 = y | V = 1, X = x)  = \dfrac{h(y, x) \exp\{\beta(y, x)\}}{\int h(y, x) \exp\{\beta(y, x)\} \;dy}
\end{align*}
where $\int h(y, x) \exp\{\beta(y, x)\} \;dy$ is a proportionality or normalizing constant, ensuring the left-hand side is a proper density. This parameterization can, in principle, represent any likelihood function one might encounter in practice and therefore imposes no restrictions on the data generation process (i.e. it is completely nonparameteric). 

For our focal outcome of testing positive, $Y = 2$, we have:
\begin{align*}
    \Pr(Y^0 = 2 | V = 1, X = x) = \dfrac{\Pr(Y^0 = 2 | V = 0, X = x)\exp\{\beta(2, x)\}}{\int f(Y^0=y | V = 0, X = x)\exp\{\beta(y, x)\} \;dy}.
\end{align*}
Because $Y$ has only three levels, the denominator may be written as
\begin{align*}
    \int & f(Y^0=y | V = 0, X = x) \exp\{\beta(y, x)\} \;dy =  \\
    &= \sum_{y=0}^2\Pr(Y^0=y| V = 0, X = x)\exp\{\beta(y, x)\} \\
    &= \dfrac{\Pr(Y^0=0| V = 0, X = x)}{\Pr(Y^0=0| V = 1, X = x)} \bigg\{\Pr(Y^0=2| V = 1, X = x) + \Pr(Y^0=1| V = 1, X = x) \\
    &\qquad   + \Pr(Y^0=0| V = 1, X = x) \bigg\} \\  
    &= \dfrac{\Pr(Y^0=0| V = 0, X = x)}{\Pr(Y^0=0| V = 1, X = x)}
\end{align*}
and therefore 
\begin{align*}
    \Pr(Y^0 = 2 | V = 1, X = x) = \Pr(Y^0 = 2 | V = 0, X = x)\exp\{\beta(2, x)\}\dfrac{\Pr(Y^0=0| V = 1, X = x)}{\Pr(Y^0=0| V = 0, X = x)}.
\end{align*}
Under Assumption \ref{app_ass3}, $\beta(2, x) = \beta(1, x)$ and thus we have: 
\begin{align*}
    \Pr(Y^0 = 2 | V = 1, X = x) = \Pr(Y^0 = 2 | V = 0, X = x)\dfrac{\Pr(Y^0 = 1 | V = 1, X = x)}{\Pr(Y^0 = 1| V = 0, X = x)},
\end{align*}
and, re-arranging, we find
\begin{align*}
    \dfrac{\Pr(Y^0 = 2| V = 1, X = x)}{\Pr(Y^0 = 2| V = 0, X = x)} = \dfrac{\Pr(Y^0 = 1 | V = 1, X = x)}{\Pr(Y^0 = 1 | V = 0, X = x)},
\end{align*}
which is expression \ref{eqn:proxy} in the main text. 

By the invariance of odds ratios 
\begin{align*}
    \beta(y, x) = \log \dfrac{\Pr(V = 1 | Y^0 = y, X = x)\Pr(V = 0 | Y^0 = 0, X = x)}{\Pr(V = 0 | Y^0 = y, X = x)\Pr(V = 1 | Y^0 = 0, X = x)}.
\end{align*}
% and therefore we have that
% \begin{align*}
%     \log \dfrac{\pi(y, x)}{1 - \pi(y, x)} = \eta(x) + \beta(y, x)
% \end{align*}
% where
% \begin{align*}
%     &\pi(y, x) = \Pr(V = 1 | Y^0 = y, X = x), \\
%     &\eta(x) = \log \dfrac{\Pr(V = 1 | Y^0 = 0, X = x)}{\Pr(V = 0 | Y^0 = 0, X = x)}.
% \end{align*}
Under Assumption \ref{app_ass3}, again $\beta(2, x) = \beta(1, x)$ implying: 
\begin{align*}
    \dfrac{\Pr(V = 1 | Y^0 = 2, X = x)}{\Pr(V = 0 | Y^0 = 2, X = x)} = \dfrac{\Pr(V = 1 | Y^0 = 1, X = x)}{\Pr(V = 0 | Y^0 = 1, X = x)},
\end{align*}
which is also discussed in the main text.

\newpage 
\section{Extended discussion of equi-confounding assumption}\label{sec:mechanisms}

    \subsection{Unmeasured health-seeking behavior}
    Among the stated motivations for the test-negative design when it was first introduced was the potential for the TND to adjust for unmeasured health-seeking behavior. Early observational studies of flu vaccines tended to overstate their effectiveness relative to randomized trials. At the time, it was hypothesized that the bias may have been caused by differential health-seeking behavior, as healthier adults were more likely to get vaccinated and also engage in other health behaviors, such as hand-washing and personal distancing, that reduce risk of infection. While hard to measure as a traditional covariate, similar health-seeking might also be revealed by an individual's propensity to pursue a test when sick and therefore conditioning on receipt of a test may effectively resolve the issue.
    
    Under our framework, if unmeasured health-seeking exerts equal influence on the risk of infection for test-positive and test-negative illness then it satisfies Assumption \ref{ass3a}, i.e.
    \[\dfrac{\Pr(I^0=2|V=1,X)}{\Pr(I^0=2|V=0,X)}=\dfrac{\Pr(I^0=1|V=1,X)}{\Pr(I^0=1|V=0,X)}.\]
    This may be more plausible when the other risk-reducing interventions that these individuals engage in are not specific to the focal illness, but instead affect risk of other infections similarly. 
    
    At the same time, if unmeasured health-seeking behavior results in equal propensity for pursuing a test when sick then it also satisfies Assumption \ref{ass3b}
    \[\dfrac{\Pr(T^0 = 1 | I^0 =2, V =1, X)}{\Pr(T^0 = 1 | I^0 =2, V =0, X)}=\dfrac{\Pr(T^0 = 1 | I^0 =1, V =1, X)}{\Pr(T^0 = 1 | I^0 =1, V =0, X)}.\]
    The plausibility of this assumption can be increased by certain design elements of the TND. Namely, the symptom screen, when well-defined and effectively implemented, equalizes the conditions under which a test is pursued while the (assumed) unavailability of testing outside health facilities ensures individuals are blind to the causative agent. 

    We note that the equi-confounding framework also allows for alternative equi-confounding structures involving unmeasured health-seeking that could lead to over- or under-optimistic estimates of vaccine effectiveness. 
    
    \subsection{Immortal time among vaccinated}
    Concerns have also been raised about the potential for design-induced biases in the test-negative design resulting from failure to explicitly emulate a randomized trial \cite{li2024comparison}. Here we show that, when the degree of design-induced bias is similar for test-positive and test-negative illnesses, they tend to cancel each other out and produce no bias in the TND estimate of vaccine effectiveness when they are equivalent. 
    
    Consider the emulation of a trial where events occurring in the first 28 days are excluded in both the vaccine arm and the control arm, as was specified in the original trials of the SARS-CoV-2 vaccines. This is often done because, biologically, it is believed that during this window the vaccine has not had time to provoke a sufficient immunological response and therefore provides no protection, although discarding these events may produce selection bias due to depletion of susceptibles when there is an effect. Regardless, attempts to emulate this result using a TND are challenged by the fact that the TND does not have a well-defined start of follow up and only the timing of vaccination and testing are known. Therefore, investigators instead discard cases where vaccination occurred less than 28 days prior to receiving a test, i.e. only among the vaccinated. This results in a form of immortal time bias as the vaccinated cannot have the outcome during this 28 day period. Assume for a second, to simplify exposition, that conditional on $X$ there is no other source of confounding. Note that under our framework, this would imply that 
    \begin{equation*}
         \frac{\Pr(Y^0 = 2 | V = 1, X)}{\Pr(Y^0 = 2 | V = 0, X)} = \delta_2(X)
    \end{equation*}
    for some $\delta_2(X) < 1$ as the person time in the first 28 days after receiving a vaccine is differentially discarded among those who received a vaccine compared to those who did not. However, if this exclusion of person time among the vaccinated is applied independently of the test result,  the bias also applies to the incidence of test-negative illness such that
    \begin{equation*}
         \frac{\Pr(Y^0 = 1 | V = 1, X)}{\Pr(Y^0 = 1 | V = 0, X)} = \delta_1(X)
    \end{equation*}
    for some $\delta_1(X) < 1$. In the special case that $\delta_2(X) = \delta_1(X)$, which could occur if the trends in the incidence of test-positive and test-negative illness among the unvaccinated are parallel over follow up (before and after the 28 day period where immortal time is accruing), we have equi-confounding and the design-induced bias is completely removed by $\Psi^*_{om}$ and $\Psi^*_{ipw}$.
    
    % \subsection{Assortativity of vaccination}

    \subsection{Correlated vaccination history}
    Prior work on the TND suggests that the design could be biased when there are other vaccines covering test-negative illness in addition to the focal vaccine for the test positive pathogen when vaccination behavior is correlated. For instance, denote by $V^*$ the vaccination against the test-negative pathogen and assume first $V^*$ is unavailable to the investigator. As shown in the DAG in eFigure \ref{fig:additional_dags} this would also violate our identifiability assumptions because it would imply the existence of a confounder that does not affect the test positive and test negative illnesses equally, i.e., 
    \[\dfrac{\Pr(I^0 = 2 | V =1, X)}{\Pr(I^0 = 2 | V =0, X)} \neq \dfrac{\Pr(I^0 = 1 | V =1, X)}{\Pr(I^0 = 1 | V =0, X)}. \]
    However, this could be resolved by measuring and adjusting for vaccination for the test negative illness. This further highlights the need to carefully consider and, where possible, document the source of the test negative illnesses. Investigators should also regularly collect full vaccination history.
    \[\dfrac{\Pr(I^0 = 2 | V =1, V^*, X)}{\Pr(I^0 = 2 | V =0, V^*, X)} = \dfrac{\Pr(I^0 = 1 | V =1,V^*,  X)}{\Pr(I^0 = 1 | V =0,V^*,  X)} \]

    \subsection{Home testing}
    Rapid diagnostic tests that provide quick confirmation of the source of a suspected infection are increasingly available for home use. The existence of these tests may invalidate Assumption \ref{ass3b} if (a) they affect the probability of seeking care when positive versus negative and (b) they are not uniformly available for all test positive and test negative illnesses. That is, under these conditions we would expect
    \[\dfrac{\Pr(T^0 = 1 | I^0 =2, V =1, X)}{\Pr(T^0 = 1 | I^0 =2, V =0, X)} \neq\dfrac{\Pr(T^0 = 1 | I^0 =1, V =1, X)}{\Pr(T^0 = 1 | I^0 =1, V =0, X)},\]
    and therefore our identifiability results in the main text would not hold. 
    
    However, let $R$ be an indicator of taking an at home rapid test (1: used rapid test, 0: did not use rapid test). Now imagine we were able to screen for use of a rapid test when recruiting for our test negative study. We may be willing to still assume 
    \[\dfrac{\Pr(T^0 = 1 | R =0, I^0 =2, V =1, X)}{\Pr(T^0 = 1 |  R =0,  I^0 =2, V =0, X)} = \dfrac{\Pr(T^0 = 1 |  R =0, I^0 =1, V =1, X)}{\Pr(T^0 = 1 |  R =0, I^0 =1, V =0, X)} \]
    In which case our identifiability results from the main text hold conditional on $R=0$. 

    Alternatively, multiplex rapid diagnostics tests that test for multiple pathogens simultaneously are increasingly available for home use. If one could limit test-negative controls to illnesses covered by multiplex testing then rapid at-home testing may affect test-seeking equally for test-positive and test-negative illnesses (unless one is perceived as more dangerous/life-threatening), in which case it may no longer bias our estimates.

    \begin{figure}[p]
    \centering
        \begin{subfigure}[t]{0.9\textwidth}
        \centering
        \begin{tikzpicture}[> = stealth, shorten > = 1pt, auto, node distance = 2.75cm, inner sep = 0pt,minimum size = 0.5pt,  very thick]
            \tikzstyle{every state}=[
              draw = none,
              fill = none
            ]
            \node[state] (x) {$X$};
            \node[state] (v) [right of=x] {$V$};
            \node[state] (i) [right of=v] {$I = 2$};
            \node[state] (t) [right of=i] {$T$};
            \node[state] (i1) [below of=i] {$I = 1$};
            \node[state] (u) [below of=x] {$U$};
            \node[state] (v1) [below of=v] {$V^*$};
   
            \path[->] (x) edge node {} (v);
            \path[->] (x) edge [out=45, in=135] node {} (i);
    
            \path[->] (v) edge node {} (i);
            
            \path[->] (i) edge [color=forestgreen] node {} (t);
            \path[->] (i1) edge [color=forestgreen] node {} (t);

            \path[->] (x) edge [out=45, in=135] node {} (t);
    
            \path[->] (v1) edge [dashed] node {} (i1);
            \path[->] (u) edge [dashed] node {} (v1);
            \path[->] (u) edge node {} (v);
            \path[->] (u) edge node {} (x);
            \path[->] (u) edge node {} (t);
            \path[->] (u) edge [color=bblue] node {} (i);
            \path[->] (u) edge [color=bblue, out=315, in=225] node  {} (i1);

        \end{tikzpicture}
        \caption{Correlated vaccination behavior}
        \end{subfigure}
        \begin{subfigure}[t]{0.9\textwidth}
        \centering
        \begin{tikzpicture}[> = stealth, shorten > = 1pt, auto, node distance = 2.75cm, inner sep = 0pt,minimum size = 0.5pt,  very thick]
            \tikzstyle{every state}=[
              draw = none,
              fill = none
            ]
            \node[state] (x) {$X$};
            \node[state] (v) [right of=x] {$V$};
            \node[state] (i) [right of=v] {$I = 2$};
            \node[state] (r) [right of=i] {$R$};
            \node[state] (t) [right of=r] {$T$};
            \node[state] (i1) [below of=i] {$I = 1$};
            \node[state] (u) [below of=v] {$U$};
   
            \path[->] (x) edge node {} (v);
            \path[->] (x) edge [out=45, in=135] node {} (i);
    
            \path[->] (v) edge node {} (i);
            
            \path[->] (i) edge [forestgreen, out=45, in=135] node {} (t);
            \path[->] (i1) edge [forestgreen] node {} (t);

            \path[->] (i) edge [forestgreen] node {} (r);
            \path[->] (i1) edge [forestgreen] node {} (r);
            \path[->] (r) edge [dashed] node {} (t);

            \path[->] (x) edge [out=45, in=135] node {} (t);
    
            \path[->] (u) edge node {} (v);
            \path[->] (u) edge node {} (x);
            \path[->] (u) edge [out=315, in=240] node {} (t);
            \path[->] (u) edge [bblue] node {} (i);
            \path[->] (u) edge [bblue] node  {} (i1);

        \end{tikzpicture}
        \caption{At home testing}
        \end{subfigure}

    \caption{Causal DAGs showing example mechanisms where the equi-confounding assumption may be invalidated or require additional refinement.}\label{fig:additional_dags}
\end{figure}
\clearpage
\newpage
    \section{Proofs of main identifiability results} \label{sec:proofs}
    For convenience, we restate below the core identifiability conditions from the main text. 
    \begin{enumerate}[label=\upshape(A\arabic*), ref=A\arabic*]
    \item\label{app_ass1} \textit{Consistency of potential outcomes}. For all individuals $i$ and for $v \in \{0, 1\}$, we have $Y_i^v = Y_i$ when $V_i = v$. 
    \item\label{app_ass2} \textit{No effect of vaccination on testing negative and symptomatic among the vaccinated}. That is, $\Pr(Y^0=1 | V = 1, X) = \Pr(Y^1=1 | V = 1, X).$
    \item\label{app_ass3} \textit{Odds ratio equi-confounding}. Degree of unmeasured confounding bias on the odds ratio scale is the same for test-positive and test-negative illnesses, i.e. 
    $$OR_2(X) = OR_1(X), $$
    $$ \text{where } OR_y(X) = \frac{\Pr(Y^0 = y | V = 1, X)\Pr(Y^0 = 0 | V = 0, X)}{\Pr(Y^0 = 0 | V = 1, X)\Pr(Y^0 = y | V = 0, X)}.$$
    \item\label{app_ass4} \textit{Overlap of vaccination among test-positives and test-negatives}. Define $\mathcal{S}_y(v)$ as the support of the law of $(Y^0 = y, V = v, X)$, then for $v$ in $\{0,1\}$, then it must be that $\mathcal{S}_2(1) \subseteq \mathcal{S}_2(0)$ and $\mathcal{S}_2(v) \subseteq \mathcal{S}_1(v).$
    %\item[(A5)] No direct effect of vaccination on test-seeking behavior among the vaccinated. That is, for $i$ in $\{1,2\}$, $\Pr[T^1 = 1 | I^1 = i, V = 1, X] = \Pr[T^0 = 1 | I^0 = i, V = 1, X].$
    \end{enumerate}

    \newpage
    \subsection{Proof of Proposition \ref{prop1}} \label{sec:proof1}
    
    \begin{proof}
    % We begin by showing that, under Assumption \ref{app_ass1} alone, the causal risk ratio for medically-attended illness among the vaccinated, $\Psi_{RRV}$, 
    % \begin{equation*}
    %     \Psi \equiv \dfrac{\Pr(Y^1=2|V=1)}{\Pr(Y^0=2|V=1)},
    % \end{equation*}
    % is equivalent to two expressions involving only the (unobserved) treatment-free potential outcome, $Y^0$, namely 
    % \begin{equation}
    %     \Psi^0_{om} \equiv \dfrac{\Pr(Y = 2 | V = 1)}{E\left[\Pr(Y = 2 | V = 0, X) \dfrac{\Pr(Y^0 = 2 | V = 1, X)}{\Pr(Y = 2 | V = 0, X)}\Big| V = 1 \right]}
    % \end{equation}
    % and 
    % \begin{equation}
    %     \Psi^0_{ipw} \equiv \dfrac{E\{V \mathbbm 1(Y = 2)\}}{E\left\{ (1 - V) \mathbbm 1(Y = 2) \dfrac{ \Pr(V = 1 | Y^0 = 2,  X)}{ \Pr(V = 0 | Y^0 = 2,  X)}\right\}}.
    % \end{equation}
    
    For the first expression, we have 
    \begin{align*}
        \Psi &= \dfrac{\Pr(Y^1=2|V=1)}{\Pr(Y^0=2|V=1)} \\
        &= \dfrac{\Pr(Y^1=2|V=1)}{E[E\{\mathbbm 1 (Y^0 = 2) | V = 1, X\} | V = 1]} \\
        &= \dfrac{\Pr(Y^1=2|V=1)}{E\left\{\int \mathbbm 1 (Y^0 = 2) f(Y^0 = y | V = 1, X) dy \mid  V = 1\right\}} \\
        &= \dfrac{\Pr(Y^1=2|V=1)}{E\left\{\int \mathbbm 1 (Y^0 = 2) \dfrac{f(Y^0 = y | V = 0, X) \exp\{\beta(y, X)\}}{\int f(Y^0 = y | V = 0, X) \exp\{\beta(y, X)\}dy}dy \Big|  V = 1\right\}} \\
        &= \dfrac{\Pr(Y^1=2|V=1)}{E\left\{ \Pr(Y^0 = 2 | V = 0, X) \dfrac{\Pr(Y^0 = 1 | V = 1, X)}{\Pr(Y^0 = 1 | V = 0, X)} \Big|  V = 1\right\}} \\
         &= \dfrac{\Pr(Y^1=2|V=1)}{E\left\{ \Pr(Y^0 = 2 | V = 0, X) \dfrac{\Pr(Y^1 = 1 | V = 1, X)}{\Pr(Y^0 = 1 | V = 0, X)} \Big| V = 1\right\}} \\
        &= \dfrac{\Pr(Y=2|V=1)}{E\left\{ \Pr(Y = 2 | V = 0, X) \dfrac{\Pr(Y = 1 | V = 1, X)}{\Pr(Y = 1 | V = 0, X)} \Big|  V = 1\right\}} \\
        &= \dfrac{\Pr(Y=2|V=1)}{E\left\{ \mathbbm 1(Y = 1) \dfrac{\Pr(Y = 2 | V = 0, X)}{\Pr(Y = 1 | V = 0, X)} \Big| V = 1\right\}} \\
        &= \dfrac{E\left\{V \mathbbm 1(Y = 2)\right\}}{E\left\{ V \mathbbm 1(Y=1) \dfrac{\Pr(Y = 2 | V = 0, X) }{\Pr(Y = 1 | V = 0, X)}  \right\}}.
    \end{align*}
    The first line restates the definition. The second uses the law of iterated expectation. The third applies the definition of conditional expectation. The fourth applies the nonparametric odds ratio model in \ref{sec:or_model}. The fifth applies the equi-confounding assumption (Assumption \ref{app_ass3}) in which $\beta(2,X) = \beta(1,X)$ and simplifies. The sixth applies the no effect on test-negative illness (Assumption \ref{app_ass2}). The seventh applies consistency (Assumption \ref{app_ass1}). The eigth reverses the iterated expectation and the ninth uses the definition of conditional expectation with the $\Pr(V=1)$ terms in numerator and denominator canceling.
    
    For the second, we have 
    \begin{align*}
        \Psi &= \dfrac{\Pr(Y^1=2|V=1)}{\Pr(Y^0=2|V=1)} \\
        &= \dfrac{E\left\{\dfrac{V}{\Pr(V = 1)} \mathbbm 1 (Y^1 = 2)\right\}}{E\left\{\dfrac{V}{\Pr(V=1)}\mathbbm 1 (Y^0 = 2)\right\}} \\
        &= \dfrac{E\{V \mathbbm 1 (Y^1 = 2)\}}{E\left[ E\{V \mathbbm 1 (Y^0 = 2)| Y^0, X\}\right]} \\
        &= \dfrac{E\{V \mathbbm 1 (Y^1 = 2)\}}{E\left\{\mathbbm 1 (Y^0 = 2) E(V | Y^0 = 2, X)\right\}} \\
        &= \dfrac{E\{V \mathbbm 1 (Y^1 = 2)\}}{E\left\{\mathbbm 1 (Y^0 = 2) \dfrac{ E(V | Y^0 = 2, X)}{ E(1-V | Y^0 = 2, X)}E(1-V|Y^0 = 2, X)\right\}} \\
        &= \dfrac{E\{V \mathbbm 1 (Y^1 =2)\}}{E\left\{(1 - V)\mathbbm 1 (Y^0 = 2) \dfrac{ \Pr(V=1 | Y^0 = 2, X)}{ \Pr(V=0 | Y^0 = 2, X)}\right\}} \\
        &= \dfrac{E\{V \mathbbm 1 (Y^1 =2)\}}{E\left\{(1 - V)\mathbbm 1 (Y^0 = 2) \dfrac{ \Pr(V=1 | Y^1 = 1, X)}{ \Pr(V=0 | Y^0 = 1, X)}\right\}} \\
        &= \dfrac{E\{V \mathbbm 1 (Y =2)\}}{E\left\{(1 - V)\mathbbm 1 (Y = 2) \dfrac{ \Pr(V=1 | Y = 1, X)}{ \Pr(V=0 | Y = 1, X)}\right\}}.
    \end{align*}
    The first line restates the definition. The second uses the definition of conditional expectation. The third applies the law of iterated expectation and cancels the $\Pr(V=1)$ terms in the numerator and denominator. The fourth multiplies by one. The fifth converts expectation to probabilities. The sixth applies the equi-confounding assumption (Assumption \ref{app_ass2}) and the no effect on test-negative illness (Assumption \ref{app_ass2}). The seventh applies consistency. 
     
    \end{proof}
    \newpage

    \subsection{Proof of Proposition \ref{prop2}}\label{sec:proof2}
    
    \begin{proof}
    Under the sampling design for a TND we define expectations as 
    \begin{align*}
        E_{TND}(Y) &= \int y f(y|S=1)dy \\
        &= E(Y|S=1)
    \end{align*}
    where $S = \mathbbm 1(Y \neq 0)$. Therefore evaluating the first expression under the sampling design we have,
        \begin{align*}
        \Psi^*_{om} &= \dfrac{E_{TND}\left\{V \mathbbm 1(Y = 2)\right\}}{E_{TND}\left\{ V \mathbbm 1(Y=1) \dfrac{E_{TND}\{\mathbbm 1 (Y = 2 )| V = 0, X\} }{E_{TND}\{\mathbbm 1 (Y = 1) | V = 0, X\}}  \right\}} \\
        &=\dfrac{E\left\{V \mathbbm 1(Y = 2) | S = 1\right\}}{E\left\{ V \mathbbm 1(Y=1)  \dfrac{E\{\mathbbm 1 (Y = 2 )| S=1, V = 0, X\} }{E\{\mathbbm 1 (Y = 1) | S= 1, V = 0, X\}}  \Big| S = 1\right\}} \\
        &=\dfrac{E\left\{\dfrac{1}{\Pr(S=1)}V \mathbbm 1(Y = 2, S = 1) \right\}}{E\left\{\dfrac{1}{\Pr(S=1)} V \mathbbm 1(Y=1, S=1)  \dfrac{E\left\{\frac{\mathbbm 1 (Y = 2, S=1 )}{\Pr(S=1|V=0,X)} \big| V = 0, X\right\} }{E\left\{\frac{\mathbbm 1 (Y = 1, S=1 )}{\Pr(S=1|V=0,X)} \big| V = 0, X\right\} }  \right\}} \\
        &=\dfrac{E\left\{V \mathbbm 1(Y = 2, S = 1) \right\}}{E\left\{ V \mathbbm 1(Y=1, S=1)  \dfrac{E\left\{\mathbbm 1 (Y = 2, S=1 ) | V = 0, X\right\} }{E\left\{\mathbbm 1 (Y = 1, S=1 ) | V = 0, X\right\}  }  \right\}} \\
        &=\dfrac{E\left\{V \mathbbm 1(Y = 2, Y\neq0) \right\}}{E\left\{ V \mathbbm 1(Y=1, Y\neq0)  \dfrac{E\left\{\mathbbm 1 (Y = 2, Y\neq0 ) | V = 0, X\right\} }{E\left\{\mathbbm 1 (Y = 1, Y\neq0 ) | V = 0, X\right\}  }  \right\}} \\
        &=\dfrac{E\left\{V \mathbbm 1(Y = 2) \right\}}{E\left\{ V \mathbbm 1(Y=1)  \dfrac{E\left\{\mathbbm 1 (Y = 2 ) | V = 0, X\right\} }{E\left\{\mathbbm 1 (Y = 1) | V = 0, X\right\}  }  \right\}} \\
        &=\dfrac{E\left\{V \mathbbm 1(Y = 2) \right\}}{E\left\{ V \mathbbm 1(Y=1)  \dfrac{\Pr(Y = 2  | V = 0, X) }{\Pr(Y = 1 | V = 0, X)  }  \right\}}.
    \end{align*}
    The first line and second lines apply the sampling design definition. The third applies the definition of conditional expectation. The fourth cancels selection probabilities $\Pr(S=1)$ and $\Pr(S=1|V=0,X)$ in numerator and denominator. The fifth uses the fact that $S = \mathbbm 1(Y \neq 0)$ for every individual under the sampling design. The sixth applies the definition of $Y$, i.e. $Y = 2 \implies Y \neq 0$. The last line convert expectations to corresponding probabilities. This shows that $\Psi^*_{om}$ is equivalent to $\Psi_{om}$ under the sampling design.
    
    For the second expression, similarly, we have,
    \begin{align*}
        \Psi^*_{ipw} &= \dfrac{E_{TND}\{V \mathbbm 1 (Y =2)\}}{E_{TND}\left\{(1 - V)\mathbbm 1 (Y = 2) \dfrac{ E_{TND}(V | Y = 1, X)}{ E_{TND}(1-V | Y = 1, X)}\right\}} \\
        &= \dfrac{E\{V \mathbbm 1 (Y =2) | S =1\}}{E\left\{(1 - V)\mathbbm 1 (Y = 2) \dfrac{ E(V | S= 1, Y = 1, X)}{ E(1-V | S = 1, Y = 1, X)}\Big| S = 1\right\}} \\
        &= \dfrac{E\left\{\dfrac{1}{\Pr(S=1)} V \mathbbm 1 (Y =2, S=1) \right\}}{E\left\{\dfrac{1}{\Pr(S=1)}(1 - V)\mathbbm 1 (Y = 2, S = 1) \dfrac{ E(V | S= 1, Y = 1, X)}{ E(1-V | S = 1, Y = 1, X)}\right\}} \\
        &= \dfrac{E\left\{ V \mathbbm 1 (Y =2, S=1) \right\}}{E\left\{(1 - V)\mathbbm 1 (Y = 2, S = 1) \dfrac{ E(V | S= 1, Y = 1, X)}{ E(1-V | S = 1, Y = 1, X)}\right\}} \\
        &= \dfrac{E\left\{ V \mathbbm 1 (Y =2, Y\neq0) \right\}}{E\left\{(1 - V)\mathbbm 1 (Y = 2, Y\neq0) \dfrac{ E(V | Y\neq0, Y = 1, X)}{ E(1-V | Y\neq0, Y = 1, X)}\right\}} \\
        &= \dfrac{E\left\{ V \mathbbm 1 (Y =2) \right\}}{E\left\{(1 - V)\mathbbm 1 (Y = 2) \dfrac{ E(V | Y = 1, X)}{ E(1-V | Y = 1, X)}\right\}} \\
        &= \dfrac{E\left\{ V \mathbbm 1 (Y =2) \right\}}{E\left\{(1 - V)\mathbbm 1 (Y = 2) \dfrac{ \Pr(V=1 | Y = 1, X)}{ \Pr(V=0 | Y = 1, X)}\right\}} .
    \end{align*}
    The first line and second lines apply the sampling design definition. The third applies the definition of conditional expectation. The fourth cancels selection probabilities $\Pr(S=1)$ in numerator and denominator. The fifth uses the fact that $S = \mathbbm 1(Y \neq 0)$ for every individual under the sampling design. The sixth applies the definition of $Y$, i.e. $Y = 2 \implies Y \neq 0$. The last line convert expectations to corresponding probabilities. This shows that $\Psi^*_{ipw}$ is equivalent to $\Psi_{ipw}$ under the sampling design.
    
    \end{proof}
    
    \newpage
    \section{Additional Estimation Details}\label{sec:app_estimation}
    \subsection{Cohort estimators}
    The identified expressions $\Psi_{om}$ and $\Psi_{ipw}$ in Proposition \ref{prop1} suggest two plug-in estimators for $\Psi$ when one has access to the full sample (including those untested) from the underlying cohort, i.e. $O_{cohort} = \{(X_i, V_i, Y_i) : i = 1, \ldots, N\}$. First, an estimator based on modeling the outcome
    \begin{equation}\label{eqn:om_estimator_cohort}
        \widehat{\Psi}_{om} = \dfrac{\sum_{i=1}^N V_i \mathbbm 1(Y_i=2)}{\sum_{i=1}^N V_i\mathbbm 1(Y_i=1) \dfrac{\widehat{\mu}_{2,0}(X_i)}{\widehat{\mu}_{1,0}(X_i)}},
    \end{equation}
    where $\mu_{y,v}(X) = \Pr(Y=y \mid V=v, X)$ is the probability of testing positive among those with vaccine status $V = v$ in the full sample which could be obtained from, for example, a multinomial logistic regression. 
    
    Second, an inverse probability weighting estimator
    \begin{equation}\label{eqn:ipw_estimator_cohort}
        \widehat{\Psi}_{ipw} = \dfrac{\sum_{i=1}^N V_i \mathbbm 1(Y_i = 2)}{\sum_{i=1}^N (1 - V_i) \mathbbm 1 (Y_i = 2) \dfrac{\widehat{\pi}_1(X_i)}{1 - \widehat{\pi}_1(X_i)}},
    \end{equation}
    where $\pi_y(X) = \Pr(V=1\mid Y=y, X)$ is the  probability of vaccination among those who test-negative. The terms $\widehat{\mu}_{1,0}(X)$, $\widehat{\mu}_{2,0}(X)$ and $\widehat{\pi}_1(X)$ are all nuisance functions. 

    \subsection{TND estimators}
    Alternatively, the identified expressions $\Psi^*_{om}$ and $\Psi^*_{ipw}$ in Proposition \ref{prop2} suggest two plug-in estimators for $\Psi$ under TND sampling, i.e. $O_{TND} = \{(X_i, V_i, S_i=1, Y^*_i) : i = 1, \ldots, n\}$. First, an estimator based on modeling the outcome
    \begin{equation}\label{eqn:om_estimator_tnd}
        \widehat{\Psi}_{om}^* = \dfrac{\sum_{i=1}^n V_i Y^*_i}{\sum_{i=1}^n V_i \widehat{\mu}^*_0(X_i)\dfrac{1 - \widehat{\mu}^*_1(X_i)}{1 - \widehat{\mu}^*_0(X_i)}},
    \end{equation}
    where $\mu^*_v(X) = \Pr(Y^*=1 \mid S=1, V=v, X)$ is the probability of testing positive among those with vaccine status $V = v$ in the full sample.

    Second, an inverse probability weighting estimator
    \begin{equation}\label{eqn:ipw_estimator_tnd}
        \widehat{\Psi}_{ipw}^* = \dfrac{\sum_{i=1}^n V_i Y^*_i}{\sum_{i=1}^n (1 - V_i) Y^*_i \dfrac{\widehat{\pi}_0^*(X_i)}{1 - \widehat{\pi}_0^*(X_i)}},
    \end{equation}
    where $\pi^*_0(X) = \Pr(V=1\mid S=1, Y^*=0, X)$ is the  probability of vaccination among those who test-negative. Both $\widehat{\mu}^*_v(X)$ and $\widehat{\pi}^*_0(X)$ are nuisance functions. 
    \newpage

    \subsection{Estimating Equations}
    Here we provide a detailed description of methods for obtaining standard errors and corresponding confidence intervals for $\widehat{\Psi}_{om}^*$ and $\widehat{\Psi}_{ipw}^*$ based on stacked estimating equations. Throughout this section we presume models for nuisance functions are correctly specified. Although not the focus of this paper, a similar procedure could be used to obtain standard errors and confidence intervals for $\widehat{\Psi}_{om}$ and $\widehat{\Psi}_{ipw}$.

    We first introduce additional notation. As before, we observe data $O = (X, V, S=1, Y^*)$ for $i = 1, \ldots, n$, where $n$ is the number of observed units. Let $\mathbb{P}(g)$ denote the average of function $g(O)$  across $n$ units, i.e., $\mathbb{P}(g) = n^{-1}\sum_{i=1}^n g(O_{i})$. Let $\operatorname{expit}(x) = \exp(x)/{1 + \exp(x)}$. For  random variables $V$ and $W$ , let $V \overset{P}{\rightarrow} W$ and $V \overset{D}{\rightarrow} W$ denote $V$ converges to $W$ in probability and in distribution, respectively.

    The estimators $\widehat{\Psi}^*_{om}$ and $\widehat{\Psi}^*_{ipw}$ can be expressed via stacked estimating equations of the form:
    \begin{equation*}
        \psi(O; \nu, \Psi^*) = \begin{pmatrix}
            \psi_{\text{nuis}}(Y^*, V, X; \nu) \\
            \psi_{\text{effect}}(Y^*, V, X; \Psi^*)
        \end{pmatrix}
    \end{equation*}
    where the first estimates the parameters $\nu$ of a model for the nuisance function $\mu_v(X)$ and the second estimates the causal estimand $\Psi^*$ for the target parameter $\Psi$, with estimates obtained as the solution to
    \begin{equation*}
        0 = \mathbb{P}\big\{\psi(O; \widehat{\nu}, \widehat{\Psi}^*)\big\}.
    \end{equation*}
    We now consider examples for each estimator in turn. 
    
    For the outcome modeling estimator, following the suggestion in the main paper, one might specify a pooled over $V$ logistic regression model for the nuisance function, i.e. $\mu^*_V(X; \nu_{om}) = \operatorname{expit}\{(1, V, X, VX)'\nu_{om}\}$. In this case, the estimating equation is given by 
    \begin{equation*}
        \psi_{om}(O; \nu_{om}, \Psi^*_{om}) = \begin{pmatrix}
            (1, V, X, VX)'\left[ Y^* - \mu^*_V(X; \nu_{om})\right] \\
            V \left\{Y^* - (1 - Y^*) \dfrac{\mu^*_0(X; \widehat{\nu}_{om})}{1 - \mu^*_0(X; \widehat{\nu}_{om})}\Psi^*_{om}\right\}
        \end{pmatrix}.
    \end{equation*}

    For the inverse probability weighting estimator, again following the suggestion in the main paper, one might specify a logistic regression model for the nuisance function, i.e. $\pi^*_0(X; \nu_{ipw}) = \operatorname{expit}\{(1, X)'\nu_{ipw}\}$. In this case, the estimating equation is given by 
    \begin{equation*}
        \psi_{ipw}(O; \nu_{ipw}, \Psi^*_{ipw}) = \begin{pmatrix}
            (1, X)'\left[V(1-Y^*) - \pi^*_0(X; \nu_{ipw})\right] \\
            Y^* \left\{V - (1-V) \dfrac{\pi^*_0(X; \widehat{\nu}_{ipw})}{1 - \pi^*_0(X; \widehat{\nu}_{ipw})}\Psi^*_{ipw}\right\}
        \end{pmatrix}.
    \end{equation*}

    We can characterize the statistical properties of the estimator using M-estimation theory \cite{stefanski_calculus_2002}. Under regularity conditions, we have the following asymptotic normality of the estimators:

    \[\sqrt{n}\left\{\big(\widehat{\Psi}^*, \widehat{\nu}\big)'-\big(\underline{\Psi}^*, \underline{\nu}\big)^\prime \right\} \xrightarrow{D} N\left(0, V_1^{-1} V_2\left(V_1^{-1}\right)^{\prime}\right)\]
    where 
    \begin{align*}
        V_1&=\left.E\left\{\frac{\partial \psi\left(O ; \Psi^*, \nu\right)}{\partial\left(\Psi^*, \nu\right)^{\prime}}\right\}\right|_{\left(\Psi^*, \nu\right)=\left(\underline{\Psi}^*, \underline{\nu}\right)}, \\
        V_2&=E\left[\left\{\Psi\left(O ;\underline{\Psi}^*, \underline{\nu}\right)\right\}\left\{\Psi\left(O ; \underline{\Psi}^*, \underline{\nu}\right)\right\}^{\prime}\right]
    \end{align*}
and where $\left(\underline{\Psi}^*, \underline{\nu}\right)$ is the probability limit of the estimators. By the law of large numbers, we have $\underline{\Psi}^*_{om}=\Psi^*_{om}$ if  $\underline{\nu}_{om}=\nu_{om}$ and $\underline{\Psi}^*_{ipw}=\Psi^*_{ipw}$ if $\underline{\nu}_{ipw}=\nu_{ipw}$, which occurs when the nuisance functions are correctly specified.

A consistent variance estimator for $\widehat{\Psi}^*$ can be obtained by substituting the empirical analogs of $V_1$ and $V_2$, i.e., $\widehat{\sigma}^2=\widehat{V}_1^{-1} \widehat{V}_2(\widehat{V}_1^{-1})^{\prime}$. Confidence intervals can be constructed leveraging the asymptotic normality of $\widehat{\Psi}^*$ and the consistent variance estimator $\widehat{\sigma}^2$. A $100(1-\alpha) \%$ confidence interval for $\widehat{\Psi}^*$ is given as
$$
\left(\widehat{\Psi}^*-z_{1-\frac{\alpha}{2}} \frac{\widehat{\sigma}}{\sqrt{n}}, \widehat{\Psi}^*+z_{1-\frac{\alpha}{2}} \frac{\widehat{\sigma}}{\sqrt{n}}\right)
$$
where $z_\alpha$ is the $100 \alpha$th percentile of the standard normal distribution.

\subsection{Bootstrapping}
An alternative method for obtaining standard errors and confidence intervals for $\widehat{\Psi}^*_{om}$ or $\widehat{\Psi}^*_{ipw}$ is the  bootstrap \cite{diciccio_bootstrap_1987}. As in the main text, let $O = \{(X_i, V_i, S_i = 1, Y_i^*)\}_{i=1}^n$ represent the available data which are  independent realizations from unknown probability distribution $F_{\theta}$. 

\begin{enumerate}
    \item Draw bootstrap replicates $O_1, \ldots, O_B$ from $\widehat{F}_{\theta}$ by sampling with replacement from the dataset. 
    \item For each replicate, $O_b$, estimate $\Psi^*$ by first estimating the nuisance function, $\mu^*_v(X)$ or $\pi_0^*(X)$, and then plug into expression \ref{eqn:om_estimator_tnd} or \ref{eqn:ipw_estimator_tnd}, thereby obtaining estimates $\widehat{\Psi}^*_1, \ldots, \widehat{\Psi}^*_B$.
    \item Calculate the bootstrapped standard error from $\widehat{\sigma}_B = \sqrt{ \sum_{i=1}^{B} (\widehat{\Psi}^*_i - \overline{\Psi}^*)^2/(B-1)}$ where $\overline{\Psi}^* = B^{-1}\sum_{i=1}^B \widehat{\Psi}^*_i$.
    \item Form a $100(1-\alpha)\%$ confidence interval for $\Psi^{\star}$ via
    $$
    \left(\widehat{\Psi}^*-z_{1-\frac{\alpha}{2}} \frac{\widehat{\sigma}}{\sqrt{n}}, \widehat{\Psi}^*+z_{1-\frac{\alpha}{2}} \frac{\widehat{\sigma}}{\sqrt{n}}\right)
    $$
    where $z_\alpha$ is the $100 \alpha$th percentile of the standard normal distribution.
\end{enumerate}
 Steps 3 and 4 may be replaced by alternative procedures which, in some cases, may yield better performing intervals.
    \newpage

     \section{Doubly robust estimator under cohort sampling}\label{sec:eif_cohort}

Recall, the parameter of interest is the causal risk ratio among the vaccinated $\Psi = \Pr(Y^1=2|V=1)/\Pr(Y^0 = 2 | V = 1)$. We focus on the denominator $\psi = \Pr(Y^0 = 2 | V = 1)$ as identification of the numerator is trivial under causal consistency. Identification of $\psi$ under equi-confounding is structurally similar to Universal Difference-in-Differences (UDiD), with a key difference that the NCO and Outcome are mutually exclusive by design. Under the UDiD framework, \textcite{tchetgen_universal_2023} provide the following estimators of $\psi$:
\begin{equation} \label{eqn:udid1}
    E\left[\dfrac{E[\mathbbm 1(Y=2)\exp\{\beta(Y,X)\} | V = 0, X]}{E[\exp\{\beta(Y,X)\} | V = 0, X]}\bigg|V = 1\right]
\end{equation}
and 
\begin{equation}\label{eqn:udid2}
    \dfrac{E[(1-V)\mathbbm 1(Y=2)\exp\{\eta(X) + \beta(Y,X)\}]}{E[(1-V)\exp\{\eta(X) +\beta(Y,X)\}]}
\end{equation}
where
\begin{align*}
    \beta(y,X) &= \log\dfrac{\Pr(Y^0=y|V=1,X)\Pr(Y^0=0|V=0,X)}{\Pr(Y^0=0|V=1,X)\Pr(Y^0=y|V=0, X)} \\
    &= \log\dfrac{\Pr(V=1 | Y^0=y,X)\Pr(V=0|Y^0=0,X)}{\Pr(V=0|Y^0=y, X)\Pr(V=1|Y^0=0,X)} \\
    \eta(X) &= \log \dfrac{\Pr(V = 1 | Y^0 = 0, X)}{\Pr(V = 0 | Y^0 = 0, X)}
\end{align*}
and $\beta(1,X) = \beta(2,X)$ under equi-confounding. \textcite{tchetgen_universal_2023} also provide a doubly-robust estimator of $\psi$, given by 
\begin{equation}\label{eqn:udid3}
     E\left[\dfrac{(1 - V)}{\Pr(V=1)} \exp\{\eta(X) + \beta(Y,X)\}\{\mathbbm 1(Y = 2) - \xi(X)\} + \dfrac{V}{\Pr(V=1)}\xi(X)  \right]
\end{equation}
where 
\[\xi(X) = \dfrac{E[\mathbbm 1(Y=2)\exp\{\beta(Y,X)\} | V = 0, X]}{E[\exp\{\beta(Y,X)\} | V = 0, X]}.\]
Equation \ref{eqn:udid3} requires correct specification of $\beta(Y, X)$ as it is common to both equation \ref{eqn:udid1} and  \ref{eqn:udid2}.  However, assuming $\beta(Y,X)$ is correctly specified, equation \ref{eqn:udid3} is consistent if either of (1) a model for the outcome under $V=0$ or (2) a model for the odds of $V$ given the reference is correct.

We begin by showing equations \ref{eqn:udid1} and \ref{eqn:udid2} are equivalent to the denominators of identifying expressions $\Psi_{om}$ and $\Psi_{ipw}$ from Proposition \ref{prop1} in the main text, i.e.
 \begin{equation*}
        E\left[\Pr[Y = 2 | V = 0, X] \dfrac{\Pr[Y = 1 | V = 1, X]}{\Pr[Y = 1 | V = 0, X]} \Big| V = 1 \right]
    \end{equation*}
    and 
    \begin{equation*}
        E\left[ \dfrac{(1 - V)}{\Pr(V=1)} \mathbbm 1(Y = 2) \dfrac{\pi(1,X)}{1 - \pi(1,X)}\right]
    \end{equation*}
where $\pi(y, X) = \Pr(V = 1 | Y^0 = y, X)$ and $\log \dfrac{\pi(Y, X)}{1 - \pi(Y, X)} = \eta(X) + \beta(Y, X)$.

We then show that equation \ref{eqn:udid3} retains its double-robustness property under our set up with
    \begin{align*}
        \xi(X) &= \dfrac{\exp\{\beta(1,X)\}}{E[\exp\{\beta(Y,X)\} | V = 0, X]}\mu_2(0,X) \\
            &= \exp\{\alpha(1,X)\}\mu_2(0,X)
    \end{align*}
    where $\alpha(y, X) = \log \dfrac{\Pr(Y^0=y|V=1,X)}{\Pr(Y^0=y|V=0,X)}$ and $\mu_y(v,X) = E[\mathbbm 1 (Y= y) | V = v, X]$. That is, Equation \ref{eqn:udid3} is consistent for $\phi$ if $\beta(Y,X)$ is correctly specified and either or both of the following hold:
    \begin{enumerate}
        \item $\mu_y^\dagger(0,X) = \mu_y(0,X)$ for $y \in \{1, 2\}$.
        \item $\eta^\dagger(X) = \eta(X)$.
    \end{enumerate}
    where  $\mu_y^\dagger(0,X)$ is a, possibly misspecified, estimator of $\mu_y(0,X)$ and likewise $\eta^\dagger(X)$ is a, possibly misspecified, estimator of $\eta(X)$.

\subsection{Proof of equivalence with UDiD}
 For equation \ref{eqn:udid1}, we have
\begin{align*}
     E&\left[\dfrac{E[\mathbbm 1(Y=2)\exp\{\beta(Y,X)\} | V = 0, X]}{E[\exp\{\beta(Y,X)\} | V = 0, X]}\bigg|V = 1\right] = \\
     &\quad = E \left[\dfrac{\int \mathbbm 1(Y=2)\exp\{\beta(y,X)\} f(y| V = 0, X)dy}{\int \exp\{\beta(y,X)\} f(y| V = 0, X)}\bigg|V = 1\right] \\
     &\quad = E \left[\dfrac{\Pr(Y = 2 | V = 0, X)\exp\{\beta(2,X)\} }{\int \exp\{\beta(y,X)\} f(y| V = 0, X)}\bigg|V = 1\right] \\
     &\quad = E \left[\dfrac{\Pr(Y = 2 | V = 0, X)\exp\{\beta(1,X)\} }{\int \exp\{\beta(y,X)\} f(y| V = 0, X)}\bigg|V = 1\right] \\
     &\quad = E \left[\dfrac{\Pr(Y = 2 | V = 0, X)\exp\{\beta(1,X)\} }{\Pr(Y^0 = 0| V = 0, X)/\Pr(Y^0 =0|V = 1, X)}\bigg|V = 1\right] \\
     &\quad = E \left[\Pr(Y = 2|V = 0, X)\dfrac{\Pr(Y = 1 | V = 1, X)}{\Pr(Y = 1 | V = 0, X)}\bigg|V = 1\right]
\end{align*}
For equation \ref{eqn:udid2}, we have
\begin{align*}
     &\dfrac{E[(1-V)\mathbbm 1(Y=2)\exp\{\eta(X) + \beta(Y,X)\}]}{E[(1-V)\exp\{\eta(X) +\beta(Y,X)\}]} = \\
     &\qquad = \dfrac{E\left[(1-V)\dfrac{\pi(2,X)}{1-\pi(2,X)}\mathbbm 1(Y=2)\right]}{E\left[(1-V)\dfrac{\pi(Y, X)}{1-\pi(Y,X)}\right]} \\
     &\qquad = \dfrac{E\left[(1-V)\dfrac{\pi(1,X)}{1-\pi(1,X)}\mathbbm 1(Y=2)\right]}{E\left[(1-V)\dfrac{\pi(Y, X)}{1-\pi(Y,X)}\right]} \\
     &\qquad = \dfrac{E\left[(1-V)\dfrac{\pi(1,X)}{1-\pi(1,X)}\mathbbm 1(Y=2)\right]}{E\left[E[(1-V) | Y, X]\dfrac{\pi(Y, X)}{1-\pi(Y,X)}\right]} \\
     &\qquad = \dfrac{E\left[(1-V)\dfrac{\pi(1,X)}{1-\pi(1,X)}\mathbbm 1(Y=2)\right]}{E\left[\pi(Y,X)\right]} \\
     &\qquad = E\left[\dfrac{(1-V)}{\Pr[V=1]}\dfrac{\pi(1,X)}{1-\pi(1,X)}\mathbbm 1(Y=2)\right]
\end{align*}
\subsection{Proof of double robustness}
When $\eta^\dagger(X) = \eta(X)$:
\begin{align*}
    E&\left[\dfrac{(1 - V)}{\Pr(V=1)} \exp\{\eta^\dagger(X) + \beta(Y,X)\}\{\mathbbm 1(Y = 2) - \xi^\dagger(X)\} + \dfrac{V}{\Pr(V=1)}\xi^\dagger(X) \right]\\
     &= E\left[\dfrac{(1 - V)}{\Pr(V=1)}  \dfrac{\pi(Y^0,X)}{1-\pi(Y^0,X)} \{\mathbbm 1(Y^0 = 2) - \xi^\dagger(X)\}  + \dfrac{V}{\Pr(V=1)}\xi^\dagger(X)\right] \\
      &= E\left[\dfrac{E[1-V|Y^0, X]}{\Pr(V=1)}  \dfrac{\pi(Y^0,X)}{1-\pi(Y^0,X)}\{\mathbbm 1(Y^0 = 2) - \xi^\dagger(X)\} + \dfrac{V}{\Pr(V=1)}\xi^\dagger(X)\right]\\
       &= E\left[\dfrac{E[V|Y^0, X]}{\Pr(V=1)} \{\mathbbm 1(Y^0 = 2) - \xi^\dagger(X)\} + \dfrac{V}{\Pr(V=1)}\xi^\dagger(X)\right] \\
       &= E\left[\dfrac{V}{\Pr(V=1)} \{\mathbbm 1(Y^0 = 2) - \xi^\dagger(X)\} + \dfrac{V}{\Pr(V=1)}\xi^\dagger(X)\right]\\
        &= E\left[\mathbbm 1(Y^0 = 2) |V=1\right]
\end{align*}
When $\mu_y^\dagger(0,X) = \mu_y(0,X) \implies \xi^\dagger(X) = \xi(X)$:
\begin{align*}
     E&\left[\dfrac{(1 - V)}{\Pr(V=1)} \exp\{\eta^\dagger(X) + \beta(Y,X)\}\{\mathbbm 1(Y = 2) - \xi^\dagger(X)\} + \dfrac{V}{\Pr(V=1)}\xi^\dagger(X) \right]\\
     &= E\left[\dfrac{(1 - V)}{\Pr(V=1)} \dfrac{\pi^\dagger(Y^0,X)}{1-\pi^\dagger(Y^0, X)}\{\mathbbm 1(Y^0 = 2) - \xi(X)\} + \dfrac{V}{\Pr(V=1)}\xi(X) \right] \\
      &= E\left[\dfrac{E[1-V|Y^0, X] }{\Pr(V=1)} \dfrac{\pi^\dagger(Y^0,X)}{1-\pi^\dagger(Y^0, X)}\{\mathbbm 1(Y^0 = 2) - \xi(X)\} + \dfrac{V}{\Pr(V=1)}\xi(X) \right] \\
      &= E\left[\dfrac{ E[V|Y^0, X]}{\Pr(V=1)}\underbrace{\dfrac{E[1-V|Y^0, X]}{E[V|Y^0, X]} \dfrac{\pi^\dagger(Y^0,X)}{1-\pi^\dagger(Y^0, X)}}_{=\exp\{(\eta^\dagger(X) - \eta(X)\}}\{\mathbbm 1(Y^0 = 2) - \xi(X)\} + \dfrac{V}{\Pr(V=1)}\xi(X) \right] \\
      &= E\left[\dfrac{ E[V|Y^0, X]}{\Pr(V=1)}\exp\{(\eta^\dagger(X) - \eta(X)\}\{\mathbbm 1(Y^0 = 2) - \xi(X)\} + \dfrac{V}{\Pr(V=1)}\xi(X) \right] \\
      &= E\left[\dfrac{V}{\Pr(V=1)}\exp\{(\eta^\dagger(X) - \eta(X)\}\{\mathbbm 1(Y^0 = 2) - \xi(X)\} + \dfrac{V}{\Pr(V=1)}\xi(X) \right] \\
      &= E\left[E\left[\exp\{(\eta^\dagger(X) - \eta(X)\}\{\mathbbm 1(Y^0 = 2) - \xi(X)\} | V = 1\right] + E\left[\xi(X)\} |V=1\right]\right] \\
      &= E\left[E\left[\exp\{(\eta^\dagger(X) - \eta(X)\}\{\underbrace{ E[\mathbbm1(Y^0 = 2)|V=1,X] - \xi(X)}_{=0}\} | V = 1\right] + E\left[\xi(X) |V=1\right]\right] \\
        &= E\left[\mathbbm1(Y^0 = 2)|V=1\right]
\end{align*}
\newpage
\subsection{Estimation}
\textbf{Steps:}
\begin{enumerate}
    \item Fit $\mu_y^\dagger(0,X)$ e.g., via multinomial regression of $Y$ on $(1,X)'$ among those with $V= 0$.
    \item Fit $\pi^\dagger(1,X)$ e.g., via logistic regression of $V$ on $(1,X)'$ among those with $Y=1$. Note that 
    \[\log \dfrac{\pi(1,X)}{1-\pi(1,X)} = \eta(X) + \beta(1, X),\]
    but as of yet we cannot differentiate between them.
    \item Estimate $\eta^\dagger(X)$ by solving 
    \[E\left[\dfrac{(1-V)}{1 -\pi(Y,X)}\right] = 1 \iff E[(1-V)\{1 - \pi^\dagger(Y,X)\} - 1] = 0\]
    where 
    \[\log \dfrac{\pi^\dagger(Y,X)}{1 - \pi^\dagger(Y,X)} = \begin{cases} \eta(X) & \text{if } Y = 0 \\
    \log \dfrac{\pi^\dagger(1,X)}{1-\pi^\dagger(1,X)}  & \text{if } Y \in \{1,2\}
    \end{cases}\]
    \item Estimate doubly-robust $\beta_{DR}^\dagger(Y,X)$ via 
    \[E\left[\{V - \operatorname{expit}(\eta^\dagger(X))\}\exp\{-\beta_{DR}(Y,X)V\}\{S(Y,X) - \mu_Y^\dagger(0, X)S(Y,X)\}\right] = 0\]
    where $S(Y,X) = \mathbbm 1 (Y = 1)$.
    \item Re-estimate $\eta_{DR}^\dagger(X)$ by solving 
    \[E\left[\dfrac{(1-V)}{1 -\pi(Y,X)}\right] = 1 \iff E[(1-V)\{1 + \exp\{\eta_{DR}^\dagger(X) + \beta^\dagger_{DR}(Y,X)\}\} - 1] = 0\]
    \item Estimate effect via
    \[\dfrac{E[V \mathbbm 1(Y =2)]}{ E\left[(1 - V) \exp\{\eta_{DR}^\dagger(X) + \beta^\dagger_{DR}(Y,X)\}\{\mathbbm 1(Y = 2) - \xi^\dagger(X)\} + V\xi^\dagger(X)\}  \right]}\]
    where
    \[\xi^\dagger(X)= \dfrac{\exp\{\beta^\dagger_{DR}(Y,X)\}}{E[\exp\{\beta^\dagger_{DR}(Y,X)\}|V=0,X ]}\mu_2^\dagger(0,X).\]
\end{enumerate}
\newpage
\section{Doubly robust estimator under TND sampling}\label{sec:eif}
Under TND sampling, i.e. $O_{TND} = \{(X_i, V_i, S_i=1, Y_i) : i = 1, \ldots, n\},$ with selection $S = \mathbbm 1(Y\neq 0)$, 
we show in the main text that equations \ref{eqn:om_estimand} and \ref{eqn:ipw_estimand}, and by extension equations \ref{eqn:udid1} and \ref{eqn:udid2}, are still identifiable by
    \begin{equation*}
        \Psi_{om}^* = \dfrac{E[V \mathbbm 1 (Y = 2)|S =1]}{E\left[V\mathbbm 1 (Y = 1) \dfrac{\Pr[Y = 2 | S = 1, V = 0, X]}{\Pr[Y = 1| S = 1, V = 0, X]}\Big| S = 1 \right]}
    \end{equation*}
    and 
     \begin{equation*}
        \Psi_{ipw}^* = \dfrac{E[V 1 (Y = 2)|S =1]}{E\left[ (1 - V) 1 (Y = 2) \dfrac{\Pr[V=1 | S=1, Y=1, X]}{\Pr[V=0 | S=1, Y=1, X]} \bigg| S = 1\right]}.
    \end{equation*}
    These equations could alternatively be parameterized as 
    \begin{equation}\label{eqn:om_estimand_tnd_alt}
        \Psi_{om}^* = \dfrac{E[V Y^*|S =1]}{E\left[V(1-Y^*) \dfrac{\mu^*(0,X)}{1 - \mu^*(0,X)}\Big| S = 1 \right]}
    \end{equation}
    and 
    \begin{equation}\label{eqn:ipw_estimand_tnd_alt}
        \Psi_{ipw}^* = \dfrac{E[V Y^*)|S =1]}{E\left[ (1 - V) Y^* \dfrac{\pi^*(0,X)}{1 - \pi^*(0,X)} \bigg| S = 1\right]}
    \end{equation}
    where $Y^* = \mathbbm 1(Y= 2)$ and $\mu^*(0,X) = \Pr(Y^*=1|S=1,V=0,X)$ and $\pi^*(0,X) = \Pr(V=1 | S=1, Y^*=0, X)$. 
    
    As shown below, both $\Psi_{om}^*$ and $\Psi_{ipw}^*$ are equivalent to
    \[\Psi^*_{dr} = \dfrac{E[V Y^*|S =1]}{E\left[VY^*\exp\{-\phi^*(X)\}\Big| S = 1 \right]}\]
    where 
    \[\phi^*(X) = \log \dfrac{\Pr[Y^*=1|S=1, V=1,X]\Pr[Y^*=0|S=1, V=1,X]}{\Pr[Y^*=0|S=1, V=0,X]\Pr[Y^*=1|S=1, V=0,X]}\]
    is the log of the conditional odds ratio function. Furthermore, as described in \textcite{tchetgen_tchetgen_doubly_2010}, a doubly-robust estimator of $\phi^*(X)$ may be obtained as the solution to the empirical analogue of the population moment equation
    \[E[(1,X)'(V-\pi^*(0,X))\exp\{-\phi^*(X)VY^*\}(Y^* - \mu^*(0,X)) | S = 1] = 0.\]
    We show this double-robustness property for $\phi^*(X)$, and by extension $\Psi^*_{dr}$, when either or both of the following conditions hold:
    \begin{enumerate}
        \item $\mu^\dagger(0,X) = \mu^*(0,X)$.
        \item $ \pi^\dagger(0,X) = \pi^*(0,X)$.
    \end{enumerate}
    where  $\mu^\dagger(0,X)$ is a, possibly misspecified, estimator of $\mu^*(0,X)$ and likewise $\pi^\dagger(0,X)$ is a, possibly misspecified, estimator of $\pi^*(0,X)$.
    \subsection{Proof of double robustness}
     Note that 
    \begin{align*}
        f(V = v | Y^*, X) &= \dfrac{\Pr(V = v | S = 1, Y^* = 0, X)\exp\{\phi^*(X)vY^*\}}{\sum_{v'}\Pr(V = v' | S = 1, Y^* = 0, X)\exp\{\phi^*(X)v'Y^*\}} \\
        &= c_v\Pr(V = v | S = 1, Y^* = 0, X)\exp\{\phi^*(X)vY^*\}
    \end{align*}
    where $c_v = 1 / \sum_{v'}\Pr(V = v' | S = 1, Y^* = 0, X)\exp\{\phi^*(X)v'Y^*\}$ is a constant. 
    \vspace{1em} \\
    \noindent Now consider $ \pi^\dagger(0,X) = \pi^*(0,X)$:
    \begin{align*}
        &E[W(X)(V-\pi^\dagger(0,X))\exp\{-\phi^*(X)VY^*\}(Y^* - \mu^\dagger(0,X)) | S = 1]  \\
        &\quad = E[W(X)(V-\pi^*(0,X))\exp\{-\phi^*(X)VY^*\}(Y^* - \mu^\dagger(0,X)) | S = 1] \\
        &\quad = E[E[W(X)(V-\pi^*(0,X))\exp\{-\phi^*(X)VY^*\}(Y^* - \mu^\dagger(0,X)) | S=1, Y^*, X] | S = 1]\\
        &\quad = E[W(X)(Y^* - \mu^\dagger(0,X))E[(V-\pi^*(0,X))\exp\{-\phi^*(X)VY^*\} | S=1, Y^*, X] | S = 1] \\
        &\quad = E[W(X)(Y^* - \mu^\dagger(0,X))\sum_v[(v-\pi^*(0,X))\exp\{-\phi^*(X)vY^*\} \Pr(V = v | S=1, Y^*, X)] | S = 1] \\
        &\quad = E[W(X)(Y^* - \mu^\dagger(0,X))c_v\sum_v\left[(v-\pi^*)\Pr(V = v | S = 1, Y^* = 0, X)\right] | S = 1] \\
        &\quad = E[W(X)(Y^* - \mu^\dagger(0,X))c_vE\left[V-\pi^*(0,X)| S = 1, Y^* = 0, X\right] | S = 1] \\
         &\quad = E[W(X)(Y^* - \mu^\dagger(0,X))c_v\underbrace{\{E\left[V| S = 1, Y^* = 0, X\right] -\pi^*(0,X)\}}_{=0}| S = 1] \\
        &\quad=0
    \end{align*}
     \vspace{1em} \\
    Similarly, note that 
    \begin{align*}
        f(Y^* = y | V, X) &= \dfrac{\Pr(Y^*=y | S = 1, V = 0, X)\exp\{\phi^*(X)Vy\}}{\sum_{y'}\Pr(Y^* = y' | S = 1, V = 0, X)\exp\{\phi^*(X)v'Y^*\}} \\
        &= c_y\Pr(Y^* = y | S = 1, V = 0, X)\exp\{\phi^*(X)Vy\}
    \end{align*}
    where $c_y = 1 / \Pr(Y^* = y' | S = 1, V = 0, X)\exp\{\phi^*(X)v'Y^*\}$ is a constant.
    \vspace{1em} \\
    Now consider $ \mu^\dagger(0,X) = \mu^*(0,X)$:
    \begin{align*}
        &E[W(X)(V-\pi^\dagger(0,X))\exp\{-\phi^*(X)VY^*\}(Y^* - \mu^\dagger(0,X)) | S = 1]  \\
        &\quad = E[W(X)(V-\pi^\dagger(0,X))\exp\{-\phi^*(X)VY^*\}(Y^* - \mu^*(0,X)) | S = 1] \\
        &\quad = E[E[W(X)(V-\pi^\dagger(0,X))\exp\{-\phi^*(X)VY^*\}(Y^* - \mu^*(0,X)) | S=1, V, X] | S = 1]\\
        &\quad = E[W(X)(V-\pi^\dagger(0,X))E[(Y^* - \mu^*(0,X))\exp\{-\phi^*(X)VY^*\} | S=1, V, X] | S = 1] \\
        &\quad = E[W(X)(V-\pi^\dagger(0,X))\sum_y[(y - \mu^*(0,X))\exp\{-\phi^*(X)Vy\} \Pr(Y^* = y | S=1, V, X)] | S = 1] \\
        &\quad = E[W(X)(V-\pi^\dagger(0,X))c_y\sum_y\left[(y - \mu^*(0,X))\Pr(Y^* = y | S=1, V=0, X)\right] | S = 1] \\
        &\quad = E[W(X)(V-\pi^\dagger(0,X))c_yE\left[Y^* - \mu^*(0,X)| S = 1, Y^* = 0, X\right] | S = 1] \\
         &\quad = E[W(X)(V-\pi^\dagger(0,X))c_y\underbrace{\{E\left[Y^*| S = 1, V = 0, X\right] -\mu^*(0,X)\}}_{=0}| S = 1] \\
        &\quad=0
    \end{align*}
    \newpage
    \subsection{Estimation}
\textbf{Steps:}
\begin{enumerate}
    \item Fit $\widehat{\mu}^*(0,X)$ e.g., via logistic regression of $Y^*$ on $(1,X)'$ among those with $V= 0$ using $O_{TND}$.
    \item Fit $\widehat{\pi}^*(0,X)$ e.g., via logistic regression of $V$ on $(1,X)'$ among those with $Y^* = 0$ using $O_{TND}$.
    \item  Estimate doubly-robust $\widehat{\phi}_{dr}^*(X)$ as solution to 
    \[\sum_{i=1}^n (1,X_i)'(V_i-\widehat{\pi}^*(0,X_i))\exp\{-\phi_{dr}^*(X_i)V_iY_i^*\}(Y^* - \widehat{\mu}^*(0,X_i)) = 0.\]
    \item Estimate the effect via
\begin{equation*}
    \widehat{\Psi}_{dr}^* = \dfrac{\sum_{i=1}^n V_i Y^*_i}{\sum_{i=1}^nV_iY^*_i \exp\{\widehat{\phi}_{dr}^*(X)\}}.
\end{equation*}
\end{enumerate}

\newpage

    \section{Direct effects of vaccination and identification of alternative estimands}\label{sec:de_testing}
    In the main text, we discuss the identification of the causal effect of vaccination on \textit{medically-attended illness} among the vaccinated, defined by the causal risk ratio:
    \[\Psi_{RRV} \equiv \dfrac{\Pr(Y^1 = 2| V = 1)}{\Pr(Y^0 = 2 | V = 1)} = \dfrac{\Pr(I^1 = 2, T^1 =1 | V = 1)}{\Pr(I^0 = 2, T^0 =1 | V = 1)}.\]
    This effect is assumed to be of substantive interest to investigators, as it is often the primary outcome in phase III vaccine trials. However, this outcome encompasses multiple component events downstream of exposure and infection, each of which may be influenced by vaccination, resulting in a non-null total effect on medically-attended illness. For example, consider a non-null effect, i.e. $\Psi_{RRV} \neq 1$ could occur if any one of the following scenarios were true:
    \begin{itemize}
        \item[(i)] The vaccine has an effect on exposure.
        \item[(ii)] The vaccine has no effect on exposure, but has an effect on infection.
        \item[(iii)] The vaccine has no effect on exposure or infection, but has an effect on the development of symptoms.
        \item[(iv)] The vaccine has no effect on exposure, infection, or symptoms, but has effect on another intermediate outcome, such as severity.
        \item[(v)] The vaccine has no effect on exposure, infection, symptoms, or severity, but has an effect on testing behavior when sick.
    \end{itemize}
    Depending on the research question, the effect on some of these components may not be clinically relevant. For instance, consider the case where the only effect of vaccination is to reduce care seeking among the vaccinated, i.e., a purely behavioral effect. This would lead to a real reduction in $\Psi_{RRV}$ even though there is no ``biological'' action of the vaccine.
    
    In this section, we explore the identification of alternative estimands that focus on specific components of the process and exclude others that are not relevant. In particular, we consider certain ``biological'' effects that exclude components where changes may stem from altered human behavior due to lack of blinding. 

    \subsection{Symptomatic infection}
    Consider first the effect of vaccination on symptomatic infection, i.e., removing the influence of any ``non-biological'' effects of unblinded vaccination on test or care-seeking behavior. Below we show that, if additional  conditions beyond Assumptions \ref{app_ass1} - \ref{app_ass4} hold, alternative estimands for the effect of vaccination on symptomatic infection are identifiable under the test-negative design. These conditions include:
    \begin{enumerate}[label=\upshape(A5\alph*), ref=A5\alph*]
        \item\label{app_ass5a}\textit{No effect of vaccination on testing for any individual, except through infection.} That is, $T^{v, i} = T^i$ for all $v$ in $\{0, 1\}$ and for all $i$ in $\{1, 2\}$. A consequence of this is 
        \[\Pr(T^1=1|I^1=1, V=1, X) = \Pr(T^0=1|I^0=1, V=1, X),\]
        and 
        \[\Pr(T^1=1|I^1=2, V=1, X) = \Pr(T^0=1|I^0=2, V=1, X).\]
         \item\label{app_ass5b}\textit{Equal effects of vaccination on testing for test-positive and test-negative illness.} That is, 
         \[\dfrac{\Pr(T^1 = 1 | I^1 = 1, V = 1)}{\Pr(T^0 = 1 | I^0 = 1, V = 1)} = \dfrac{\Pr(T^1 = 1 | I^1 = 2, V = 1)}{\Pr(T^0 = 1 | I^0 = 2, V = 1)}\]
    \end{enumerate}
    \begin{enumerate}[label=\upshape(A\arabic*), ref=A\arabic*, start=6]
         \item\label{app_ass6}\textit{No effect modification for symptomatic infection among the tested on the ratio scale after conditioning on $X$} That is, 
          \[\dfrac{\Pr(I^1 = 2 | T = 1, V = 1, X)}{\Pr(I^0 = 2 | T = 1, V = 1, X)} = \dfrac{\Pr(I^1 = 2 | V = 1, X)}{\Pr(I^0 = 2 | V = 1, X)}\]
    \end{enumerate}
    The first and second conditions, \ref{app_ass5a} and \ref{app_ass5b}, are alternatives that place limits on the nature of effects on testing behavior. Assumption \ref{app_ass5a} requires no effect of vaccination on testing behavior regardless of infection source. It might be plausible if, as in phase III efficacy studies, participants were blind to whether they received vaccination or a placebo. However, in ``real world'' settings typical of test-negative studies, \ref{app_ass5b} may be more realistic. It only requires the effects be equal regardless of whether the source of symptomatic infection is the target pathogen or alternative test-negative pathogens. This may be plausible when participants do not have access to home testing and when they are unaware of, e.g. the testing status of others in their household or contact networks who may have recently been infected. Finally, Assumption \ref{app_ass6} requires no effect modification on the relative scale between the tested sample and the target population, that is either the covariates in $X$ include all possible modifiers or the distribution of unmeasured effect modifiers is equal between the tested sample and the target population. 

    As we show in Proposition \ref{prop3} under \ref{app_ass1} - \ref{app_ass4} and either of \ref{app_ass5a} or \ref{app_ass5b} the conditional risk ratio for symptomatic infection among the vaccinated is identified by the conditional odds ratio under test-negative sampling. Likewise, in Proposition \ref{prop4}, we show that the risk ratio for symptomatic infection among the vaccinated and tested is identified by $\Psi^*_{om}$ and $\Psi^*_{ipw}$, i.e., the novel TND estimands introduced in the main text.
    \newpage
    
    \begin{proposition}\label{prop3}
    The conditional risk ratio for symptomatic infection among the vaccinated, 
    \[\Theta_{RRV}(X) \equiv \dfrac{\Pr(I^1 = 2 | V = 1, X)}{\Pr(I^0 = 2| V = 1, X)},\]
    is identified by the conditional odds ratio, 
    \begin{equation}
    \Psi^*_{om}(X) = \dfrac{\Pr(Y = 2 | S = 1, V = 1, X)/\Pr(Y = 1 | S = 1, V = 1, X)}{\Pr(Y = 2 | S = 1, V = 0, X)/\Pr(Y = 1 | S = 1, V = 0, X)},
    \end{equation}  
    under test-negative sampling provided that \ref{app_ass1} - \ref{app_ass4} hold and either of \ref{app_ass5a} or \ref{app_ass5b} hold.
    
    \end{proposition}
    \begin{proof}
        Under Assumption \ref{app_ass3}, we have that 
        \[\dfrac{\Pr(T^0 = 1 | I^0 = 2, V = 1, X)}{\Pr(T^0 = 1 | I^0 = 1, V = 1, X)} = \dfrac{\Pr(T^0 = 1 | I^0 = 2, V = 0, X)}{\Pr(T^0 = 1 | I^0 = 1, V = 0, X)}.\]
        Likewise, Assumptions \ref{app_ass5a} and \ref{app_ass5b} both imply that 
        \[\dfrac{\Pr(T^0 = 1 | I^0 = 2, V = 1, X)}{\Pr(T^0 = 1 | I^0 = 1, V = 1, X)} = \dfrac{\Pr(T^1 = 1 | I^1 = 2, V = 1, X)}{\Pr(T^1 = 1 | I^1 = 1, V = 1, X)}.\]
        Substituting and applying consistency, this implies the following restriction on the observed data
        \[\dfrac{\Pr(T = 1 | I = 2, V = 1, X)}{\Pr(T = 1 | I = 2, V = 1, X)}= \dfrac{\Pr(T = 1 | I = 2, V = 0, X)}{\Pr(T = 1 | I = 1, V = 0, X)}\]        
        Now, we can show that
        \begin{align*}
            \Theta&_{RRV}(X) = \\
            &= \dfrac{\Pr(I^1 = 2 | V = 1, X)}{\Pr(I^0 = 2| V = 1, X)} \\
            &= \dfrac{\Pr(I^1 = 2 | V = 1, X)}{\dfrac{\Pr(I^0 = 1| V = 1, X)}{\Pr(I^0 = 1| V = 0, X)}\Pr(I^0 = 2| V = 0, X)} \\
            &= \dfrac{\Pr(I^1 = 2 | V = 1, X)}{\dfrac{\Pr(I^1 = 1| V = 1, X)}{\Pr(I^0 = 1| V = 0, X)}\Pr(I^0 = 2| V = 0, X)} \\
            &= \dfrac{\Pr(I = 2 | V = 1, X)\Pr(I = 1| V = 0, X)}{\Pr(I = 1| V = 1, X)\Pr(I = 2| V = 0, X)} \\
            &= \dfrac{\Pr(I = 2 | V = 1, X)\Pr(I = 1| V = 0, X)}{\Pr(I = 1| V = 1, X)\Pr(I = 2| V = 0, X)} \times \dfrac{\Pr(T = 1 | I = 2, V = 1)\Pr(T = 1 | I = 1, V = 0)}{\Pr(T = 1 | I = 1, V = 1)\Pr(T = 1 | I = 2, V = 0)} \\
            &= \dfrac{\Pr(I = 2 | T=1, V = 1, X)\Pr(I = 1| T=1, V = 0, X)}{\Pr(I = 1| T=1, V = 1, X)\Pr(I = 2| T = 1, V = 0, X)}  \\
             &= \dfrac{\Pr(Y = 2 | S=1, V = 1, X)\Pr(Y = 1| S=1, V = 0, X)}{\Pr(Y = 1| S=1, V = 1, X)\Pr(Y = 2| S = 1, V = 0, X)}  
        \end{align*}
        where the first line follows by definition, the second applies Assumption \ref{ass3a}, the third applies Asssumption \ref{app_ass2}, the fourth applies consistency, the fifth applies the result above, the sixth uses Bayes' rule, and the last applies the definitions of $Y$ and $S$.
    \end{proof}

    \newpage
    \begin{proposition}\label{prop4}
    The risk ratio for symptomatic infection among the vaccinated and tested, 
    \[\Theta_{RRVT} \equiv \dfrac{\Pr(I^1 = 2 | T = 1, V = 1)}{\Pr(I^0 = 2| T = 1, V = 1)},\]
    is identified under test-negative sampling by
   \begin{equation*}
        \Psi_{om}^* = \dfrac{\Pr(Y = 2 | S = 1, V = 1)}{E\left\{ \Pr(Y = 2 | S = 1, V = 0, X) \dfrac{\Pr(Y = 1 | S = 1, V = 1, X)}{\Pr(Y = 1| S = 1, V = 0, X)}\Big| S = 1, V = 1 \right\}}
    \end{equation*}
    and 
    \begin{equation*}
        \Psi_{ipw}^* = \dfrac{E\left\{V \mathbbm 1 (Y = 2)|S =1\right\}}{\E\left\{ (1 - V) \mathbbm 1 (Y = 2) \dfrac{\pi^*_1(X)}{1 - \pi^*_1(X)} \bigg| S = 1\right\}}
    \end{equation*}
    where $\pi^*_1(X) = \Pr(V = 1| S = 1, Y = 1, X)$, provided that assumptions \ref{app_ass1} - \ref{app_ass4} and \ref{app_ass6} hold and either of \ref{app_ass5a} or \ref{app_ass5b} hold.
    \end{proposition}
    \begin{proof}
        Applying same arguments in Proposition \ref{prop3}, we have that 
        \[\dfrac{\Pr(T = 1 | I = 2, V = 1, X)}{\Pr(T = 1 | I = 2, V = 1, X)}= \dfrac{\Pr(T = 1 | I = 2, V = 0, X)}{\Pr(T = 1 | I = 1, V = 0, X)}\]        
        Now, we can show that
        \begin{align*}
            \Theta&_{RRVT} \\
            &= \dfrac{\Pr(I^1 = 2 | T = 1, V = 1)}{\Pr(I^0 = 2| T = 1, V = 1)} \\
            &= \dfrac{\Pr(I^1 = 2 | T = 1, V = 1)}{E\left\{\Pr(I^0 = 2| T = 1, V = 1, X)\Big| T = 1, V = 1\right\}} \\
            &= \dfrac{\Pr(I^1 = 2 | T = 1, V = 1)}{E\left\{\dfrac{\Pr(I^0 = 2 | V = 1, X)}{\Pr(I^1 = 2 | V = 1, X)}\Pr(I^1 = 2| T = 1, V = 1, X)\Big| T = 1, V = 1\right\}} \\
            &= \dfrac{\Pr(I^1 = 2 | T = 1, V = 1)}{E\left\{\dfrac{\Pr(I^0 = 1| V = 1, X)}{\Pr(I^0 = 1| V = 0, X)}\dfrac{\Pr(I^0 = 2 | V = 0, X)}{\Pr(I^1 = 2 | V = 1, X)}\Pr(I^1 = 2| T = 1, V = 1, X)\Big| T = 1, V = 1\right\}} \\
            &= \dfrac{\Pr(I^1 = 2 | T = 1, V = 1)}{E\left\{\dfrac{\Pr(I^1 = 1| V = 1, X)}{\Pr(I^0 = 1| V = 0, X)}\dfrac{\Pr(I^0 = 2 | V = 0, X)}{\Pr(I^1 = 2 | V = 1, X)}\Pr(I^1 = 2| T = 1, V = 1, X)\Big| T = 1, V = 1\right\}} \\
            &= \dfrac{\Pr(I = 2 | T = 1, V = 1)}{E\left\{\dfrac{\Pr(I = 1| V = 1, X)}{\Pr(I = 1| V = 0, X)}\dfrac{\Pr(I = 2 | V = 0, X)}{\Pr(I = 2 | V = 1, X)}\Pr(I = 2| T = 1, V = 1, X)\Big| T = 1, V = 1\right\}} \\
            &= \dfrac{\Pr(I = 2 | T = 1, V = 1)}{E\left\{\dfrac{\Pr(I = 1|T=1, V = 1, X)}{\Pr(I = 1|T = 1, V = 0, X)}\dfrac{\Pr(I = 2 | T = 1, V = 0, X)}{\Pr(I = 2 | T = 1, V = 1, X)}\Pr(I = 2| T = 1, V = 1, X)\Big| T = 1, V = 1\right\}} \\
            &= \dfrac{\Pr(I = 2 | T = 1, V = 1)}{E\left\{\dfrac{\Pr(I = 1|T=1, V = 1, X)}{\Pr(I = 1|T = 1, V = 0, X)}\Pr(I = 2 | T = 1, V = 0, X)\Big| T = 1, V = 1\right\}} \\
            &= \dfrac{\Pr(Y = 2 | S = 1, V = 1)}{E\left\{ \Pr(Y = 2 | S = 1, V = 0, X) \dfrac{\Pr(Y = 1 | S = 1, V = 1, X)}{\Pr(Y = 1| S = 1, V = 0, X)}\Big| S = 1, V = 1 \right\}}
        \end{align*}
        where the first line follows by definition, the second applies Assumption \ref{ass3a}, the third applies Asssumption \ref{app_ass2}, the fourth applies consistency, the fifth applies the result above, the sixth uses Bayes' rule, and the last applies the definitions of $Y$ and $S$.
        
    \end{proof}
    \newpage

   %  \subsection{Effect among exposed}

   %  We assume that exposure is necessary for infection and infection is necessary for the development of symptoms or severe disease which can be formalized as:
   %  \begin{itemize}
   %      \item [(A5)] \textbf{Exposure necessity for infection}. That is, $E = 0 \implies I = 0$ and $I^{e=0} = I$ if $E = 0$.
   %      \item [(A6)] \textbf{Infection necessity for symptoms and severe sequelae}. That is, \\$I = 0 \implies (W = 0, Z = 0)$ and $W^{i=0} = W$ if $W = 0$ and $Z^i=Z$ if $Z=0$.
   %  \end{itemize}
   %  We also assume throughout
   %  \begin{itemize}
   %      \item [(A7)] \textbf{No effect of vaccination on exposure}. That is, $E^{v} = E$ for all $v$ in $\{0, 1\}$.
   %      \item [(A7*)] \textbf{Equal effects of vaccination on exposure}. That is, 
   %      \[\dfrac{\Pr[E^1=2 | V=1]}{\Pr[E^0=2 | V=1]} = \dfrac{\Pr[E^1=1 | V=1]}{\Pr[E^0=1 | V=1]}\]
   %  \end{itemize}

   %  \subsection{Effect on severe sequelae}

   %  \[\Phi_{RRV} \equiv \dfrac{\Pr[I^1 = 2 | V = 1]}{\Pr[I^0 = 2 | V = 1]}.\]

   % $k$ indexes the source (1: test-negative pathogen, 2: test-positive pathogen):

   %  We can then re-create the DAG in Figure \ref{fig:dag_split} 

    \newpage
    
\clearpage

    \newpage

\newpage 

\section{Identification when test is imperfect}\label{sec:testing}
In the main text, we assumed the availability of a perfect test such that $\widetilde{Y}_i = Y_i$ for all individuals where $\widetilde{Y}$ is the test result and $Y$ is the true value. In the absence of such a test, here we describe the potential for bias due to misclassification of case status which has also been described previously \cite{sullivan_theoretical_2016}. Because most studies employ real-time reverse-transcription polymerase chain reaction (RT-PCR), false positives are unlikely as most RT-PCR tests have specificity approaching 100\%. Therefore, misclassification of test-positives is probably rare, perhaps due to sample contamination or data entry errors. False negatives may be more likely as test sensitivity is generally lower due to the fact that (1) some of those infected with test-positive illness may not shed detectable virus or viral RNA; (2) some may seek care after shedding has ceased; or (3) sample quality may be compromised due to swab quality or inadequate storage. If sample collection is sufficient such that the source of the test-negative infection can be identified via RT-PCR, then sensitivity with respect to the test-positive illness can be improved by limiting to test-negative controls with an identified cause. 

In the ideal case that we have a test with perfect specificity and there is no effect of vaccination on the measurement error, i.e. when
 \begin{enumerate}[label=\upshape(G\arabic*), ref=G\arabic*]
        \item\label{app_assg1}\textit{No false-positives (perfect specificity).} This implies that for the true and tested values, $Y$ and $\widetilde{Y}$, and for every individual $i$, we have
        \[Y_i \neq 2 \implies \widetilde{Y}_i \neq 2.\]
        \item\label{app_assg2}\textit{No effect of vaccination on test result} That is, for any $v$ and $v'$, we have
         \[\Pr(\widetilde{Y}^{1} = 2 | V = 1, Y^1=2) = \Pr(\widetilde{Y}^{0} = 2 | V = 1, Y^{0}=2).\]
    \end{enumerate}
we can actually show that the causal risk ratio using the mismeasured outcome $Y^*$ is the same as the one we would obtain with the true outcome $Y$. 
\begin{proof}
    \begin{align*}
         \dfrac{\Pr(\widetilde{Y}^{1} = 2 | V = 1)}{\Pr(\widetilde{Y}^0 = 2 | V = 1)} &= \dfrac{\sum_y \Pr(\widetilde{Y}^1 = 2, Y^1 = y | V = 1)}{\sum_y \Pr(\widetilde{Y}^0 = 2, Y^0 = y| V = 1)} \\
        &= \dfrac{\Pr(\widetilde{Y}^1 = 2, Y^1 = 2 | V = 1) + \Pr(\widetilde{Y}^1 = 2, Y^1 \neq 2 | V = 1)}{\Pr(\widetilde{Y}^0 = 2, Y^0 = 2 | V = 1) + \Pr(\widetilde{Y}^0 = 2, Y^0 \neq 2 | V = 1)} \\
        &= \dfrac{\Pr(\widetilde{Y}^1 = 2, Y^1 = 2 | V = 1)}{\Pr(\widetilde{Y}^0 = 2, Y^0 = 2 | V = 1)} \\
        &= \dfrac{\Pr(\widetilde{Y}^1 = 2 | V = 1, Y^1 = 2)\Pr(Y^1 = 2 | V = 1) }{\Pr(\widetilde{Y}^0 = 2 | V = 1, Y^0 = 2) \Pr(Y^0 = 2 | V = 1)} \\
         &= \dfrac{\Pr(Y^1 = 2 | V = 1) }{\Pr(Y^0 = 2 | V = 1) }
    \end{align*}
    The first line applies the law of total probability, the second expands the sum, the third applies \ref{app_assg1}, the fourth factors the joint probability, and the last applies \ref{app_assg2}.
\end{proof}

\newpage

\section{Simulation details}\label{sec:moresim}
Here, we provide additional details on the data generation process and estimation methods for the simulation study described in the main text. We also present additional results. 

Our data generation process is a special case of the nonparametric structural equation model implied by the single world intervention graph (SWIG) in eFigure \ref{fig:simswig} representing the data as observed in the test negative design under a hypothetical intervention that sets $V$ to $v$. More specifically, we generate data according to  
\begin{align*}
    X, U &\sim \text{Unif}(0,1)\\
    V\mid X, U & \sim \text{Bernoulli}(\expit(\alpha_0 + \alpha_X X + \alpha_U U))\\
    I^v \mid V, X, U &\sim \text{Multinomial}(1-p_1(v, X, U) - p_2(v, X, U), p_1(v, X, U), p_2(v, X, U))\\
    T^v\mid I^v=i, V, X, U &\sim \text{Bernoulli}(\mathbbm 1(i>0)\exp\{(\tau_{1} + \tau_{1V}v) \mathbbm 1(i=1) + (\tau_{2} + \tau_{2V} v + \tau_{2U} U ) \mathbbm 1(i=2) \\&\qquad \qquad\qquad + \tau_X X + \tau_U U \})
\end{align*}
where $U$ is an unmeasured confounder and
\begin{align*}
    p_1(v, X, U) & = \Pr(I^v = 1 | V, X, U) = \exp(\beta_{10} + \beta_{1V}v + \beta_{1X}X + \beta_{1VX}vX + \beta_{1U}U) \\
    p_2(v, X, U) & = \Pr(I^v = 2 | V, X, U) = \exp(\beta_{20} + \beta_{2V}v + \beta_{2X}X + \beta_{2VX}vX + \beta_{2U}U).
\end{align*}
This implies the conditional independence $Y^v \indep V\mid X, U$ holds for $v=0,1$, meaning that conditioning on $U$ and $X$ is sufficient to control confounding. Under this setup, the causal risk ratio for medically-attended illness is 
\begin{align*}
    \Psi &= \dfrac{\Pr(Y^1=2 \mid V=1)}{\Pr(Y^0=2\mid V=1)}\\
    &= \dfrac{E\{\Pr(Y^1=2 \mid V=1, U, X)\mid V=1\}}{E\{\Pr(Y^0=2\mid V=1, U, X)\mid V=1\}}\\
   % &= \dfrac{E\{\exp(\beta_{20} + \beta_{2V} + \beta_{2X}X + \beta_{2VX}X + \beta_{2U}U + \tau_2 + \tau_{2V} + \tau_X X + \tau_U U + \tau_{2U} U)\mid V=1\}}{E\{\exp(\beta_{20} +  \beta_{2X}X  + \beta_{2U}U + \tau_2 + \tau_X X + \tau_U U + \tau_{2U} U)\mid V=1\}}\\
    &= \exp(\beta_{2V} + \tau_{2V}) \dfrac{E[\exp\{(\beta_{2X} + \beta_{2VX} + \tau_X) X + (\beta_{2U} + \tau_U + \tau_{2U}) U\}\mid V=1]}{E[\exp\{ (\beta_{2X} + \tau_X)X  + (\beta_{2U} + \tau_U + \tau_{2U}) U\}\mid V=1]}
    % &= \exp(\beta_{2V} + \tau_{2V}) \dfrac{E[\exp\{(\beta_{2X} + \beta_{2VX} + \tau_X) X\}\mid V=1]}{E[\exp\{ (\beta_{2X} + \tau_X)X\}\mid V=1]}.
\end{align*}
When there is no effect modification by X ($\beta_{2VX} = 0$), this reduces to 
\begin{align*}
    \Psi &= \exp(\beta_{2V} + \tau_{2V}),
\end{align*}
and when there is no no direct effect of vaccination on testing ($\tau_{2V} = 0$), this simplifies further to $\exp(\beta_{2V})$ which we note is the same as the effect on symptomatic infection. Finally, under TND sampling with $S=\mathbbm{1}(Y \neq 0)$, it can also be shown that
\begin{align*}
    \Pr[Y=2 \mid  S=1, V, X, U] &= \expit\{(\beta_{20}-\beta_{10}+\tau_2 - \tau_1) + (\beta_{2V} - \beta_{1V} + \tau_{2V} - \tau_{1V}) V\\ &\qquad\qquad+(\beta_{2X}-\beta_{1X})X+ (\beta_{2VX} - \beta_{1VX})VX + (\beta_{2U} - \beta_{1U} + \tau_{2U})U\},
\end{align*}
which follows a logistic regression model.

Violations of Assumption \ref{ass3} are controlled via $\beta_{2U}$, $\beta_{1U}$ and $\tau_{2U}$. Under our set up, we have:
\begin{align*}
    \dfrac{\Pr[Y^0=2 \mid V=1,X]}{\Pr[Y^0=2\mid V=0,X]}=\dfrac{\E[\exp\{(\tau_U  + \tau_{2U}+\beta_{2U})U\}\mid V=1, X]}{\E[\exp\{(\tau_U  + \tau_{2U}+\beta_{2U})U\}\mid V=0, X]},
\end{align*}
and 
\begin{align*}
    \dfrac{\Pr[Y^0=1 \mid V=1,X]}{\Pr[Y^0=1\mid V=0,X]}=\dfrac{\E[\exp\{(\tau_U  +\beta_{1U})U\}\mid V=1, X]}{\E[\exp\{(\tau_U +\beta_{1U})U\}\mid V=0, X]},
\end{align*}
implying that that Assumption \ref{ass3} holds when $\beta_{2U}=\beta_{1U}$ and $\tau_{2U}=0$ (Note that under these conditions the logistic model above similarly does not depend on $U$). Following the previous discussion, Assumption \ref{ass3} can be viewed as composed of two sub conditions (\ref{ass3a} and \ref{ass3b}): First, the unmeasured confounder exerts an equivalent effect on $I=1$ and $I=2$ (i.e. $\beta_{1U}=\beta_{2U}$), and second, there is no interaction between the unmeasured confounder and the source of illness for the probability of testing on the multiplicative scale (i.e. $\tau_{2U}=0$). 

Likewise, violations of Assumption \ref{ass2} are controlled by $\beta_{1V}$ and $\tau_{1V}$. When $\beta_{1V} \neq 0$ there is a direct effect of vaccination on test-negative illness $I = 1$ and when $\tau_{1V} \neq 0$ there is a direct effect of vaccination on testing among those with test-negative illness. When this occurs, we can at best identify the ratio of vaccine effects on the two illness outcomes, i.e. $\exp(\beta_{2V} + \tau_{2V})/\exp(\beta_{1V} + \tau_{1V})$, provided there is no effect modification by $X$ or $U$. As we show in Appendix \ref{sec:de_testing} and illustrate further via simulation below, under the special case of equal effects on testing, i.e. $\tau_{2V} = \tau_{1V}$, certain effects on symptomatic illness may still be recovered.

\begin{figure}[t]
    \centering
    \begin{tikzpicture}
        
        \tikzset{line width=1.25pt,
            swig vsplit={inner line width right = 0.5pt, line width right = 2pt},
            ell/.style={draw, inner sep=3pt,line width=1.25pt}}
        
        \node[name=V, shape=swig vsplit]{ \nodepart{left}{$V$} \nodepart{right}{$v$} };

        \node (Vname) at (0, 0.5-1.75) {Vaccination};

        \node[shape=ellipse,ell] (I) at (2.5,0) {$I^v$};
        \node (Iname) at (2.5,0.5-1.75) {Symptomatic};
        \node (Iname2) at (2.5,0.5-2.25) {Illness};

        \node[shape=ellipse,ell] (T) at (5,0) {$T^v$};
        \node (Tname) at (5,0.5-1.75) {Testing};

        \node[shape=ellipse,ell] (U) at (2.5,1.5+1) {$U,X$};
        %			\node (HSname) at (0,1.5+0.5+0.5) {Unmeasured};
        \node (Uname) at (2.5,1.5+0.5+0.75+1) {Measured \& unmeasured};
        \node (Uname2) at (2.5,1.5+0.75+1) {confounding};
        \draw[-stealth, line width = 1.25pt] (U) to (-0.2, 0.31);
            \draw[-stealth, line width=1.25pt, bend right](V) to (T);

        \foreach \from/\to in {V/I, I/T, U/I, U/T}
        \draw[-stealth, line width = 1.25pt] (\from) -- (\to);
        %%%NCs

    \end{tikzpicture}
    \caption{A Single-World Intervention Graph (SWIG) for causal relationship between variables of a test-negative design in the simulation.\label{fig:simswig}}
\end{figure} 

 We generate a target population of $N = 15,000$ resulting in TND samples of test-positive cases and test-negative controls of between $2000$ and $3000$. We consider nine scenarios: 
 \begin{enumerate}
    \item No unmeasured confounding.
    \item Unmeasured confounding, but assumptions \ref{ass1}-\ref{ass4} hold.
    \item Direct effect of vaccination on $I=1$.
    \item Equi-confounding is violated.
    \item Equi-selection is violated.
    \item Equal effects of vaccination on testing.
    \item Unequal effects of vaccination on testing.
    %\item Infections $I=1$ and $I=2$ are not mutually exclusive.
    \item All assumptions hold but there is effect modification by covariates $X$.
 \end{enumerate}
 Parameter values for all scenarios are shown in eTable \ref{tab:simparams}. For each scenario, we generate 1000 replicates and estimate $\Psi$ using the proposed estimators $\widehat{\Psi}_{om}^*$, $\widehat{\Psi}_{ipw}^*$, and $\widehat{\Psi}_{dr}^*$ as well as the conventional logistic regression estimator and calculate the bias and coverage of 95\% confidence intervals based on the true value. For comparison, we also estimate $\Psi$  assuming one had access to the full sample, as in cohort study, using the following estimators
 \begin{equation}\label{eqn:cohort_estimator}
    \widehat{\Psi}_{cohort} = \dfrac{\sum_{i=1}^N\widehat{\mu}_1(X_i)}{\sum_{i=1}^N\widehat{\mu}_0(X_i)}
 \end{equation}
 and 
 \begin{equation}\label{eqn:cohort_u_estimator}
    \widehat{\Psi}_{cohort,U} = \dfrac{\sum_{i=1}^N\widehat{\mu}_1(X_i, U_i)}{\sum_{i=1}^N\widehat{\mu}_0(X_i, U_i)}
 \end{equation}
 where $\mu_v(X) = \Pr(Y=2 | X, V=v)$ and $\mu_v(X,U) = \Pr(Y=2 | X, U, V=v)$
 The first estimator represents the typical case that the covariate $U$ is unmeasured, and the second represents the hypothetical case where $U$ is available, allowing for conditioning on a sufficient adjustment set. We consider both the typical case that the covariate $U$ is unmeasured, and the hypothetical case where $U$ is available, allowing for conditioning on a sufficient adjustment set. Finally, we include estimates from and $\widehat{\Psi}_{om}$ based on $\Psi_{om}$ from Proposition \ref{prop1} and discussed further in \ref{sec:app_estimation} that uses the full sample instead of restricting to those tested. Finally, in scenario 8, we demonstrate the double robustness property of $\widehat{\Psi}_{dr}^*$ by adjusting the data generation process for $V$ to be
 \begin{align*}
    V\mid X, U & \sim \text{Bernoulli}(\exp\{\alpha_0 + \alpha_X \mathbbm 1 (X > 0.5) + \alpha_U U\})
\end{align*}
when simulating under misspecification of the extended propensity score model and adjusting the process for $I^v$ and $T^v$ to be
\begin{align*}
    \Pr(I^v = 2 | V, X, U) = \exp(\beta_{20} + \beta_{2V}v + \beta_{2X}(X - 0.5)^2 + \beta_{2VX}vX + \beta_{2U}U). \\
    \Pr(T^v = 1 | I^v = 2, V, X, U) = \exp(\beta_{20} + \beta_{2V}v + \beta_{2X}(X - 0.5)^2 + \beta_{2VX}vX + \beta_{2U}U). \\
\end{align*}
For that last scenario, we also increase the number of Monte Carlo simulations to 2000. More details on naming convention and model specification for each estimation method are provided in eTable \ref{tab:methods}. For all estimators other than TND, we estimate 95\% confidence intervals using stacked estimating equations.

\begin{table}[p]
    \centering
    \caption{Parameter values for the data generation processes in each scenario for the simulation study. Parameter values that change across simulations are highlighted in \textbf{bold}.}\label{tab:simparams}
    \begin{tabular}{ccccccccc}
        \toprule
        & \multicolumn{8}{c}{Scenario} \\
        \cmidrule{2-9}
        Parameter & 1 & 2 & 3 & 4 & 5 & 6 & 7 & 8  \\
        \midrule
        $\alpha_0$ & -0.9 & -0.9 & -0.9 & -0.9 & -0.9 & -0.9 & -0.9 & -0.9 \\
        $\alpha_X$ & -1 & -1 & -1 & -1 & -1 & -1 & -1 & -1 \\
        $\alpha_U$ & 0 & \textbf{2} & \textbf{2} & \textbf{2} & \textbf{2} & \textbf{2} & \textbf{2} & \textbf{2} \\
        $\beta_{10}$ & -2.1 & -2.1 & -2.1 & -2.1 & -2.1 & -2.1 & -2.1 &  -2.1 \\
        $\beta_{1V}$ & 0 & 0 & \textbf{0.1} & 0 & 0 & 0 & 0 & 0 \\
        $\beta_{1X}$ & -0.5 & -0.5 & -0.5 & -0.5 & -0.5 & -0.5 & -0.5 & -0.5 \\
        $\beta_{1VX}$ & 0 & 0 & 0 & 0 & 0 & 0 & 0 & 0 \\
        $\beta_{1U}$ & 1 & 1 & 1 & \textbf{0.25} & 1 & 1 & 1 & 1 \\
        $\beta_{20}$ & -2.4 & -2.4 & -2.4 & -2.4 & -2.4 & -2.4 & -2.4 &  -2.4 \\
        $\beta_{2V}$ & -1 & -1 & -1 & -1 & -1 & -1 & -1 &  \textbf{-0.25} \\
        $\beta_{2X}$ & -0.625 & -0.625 & -0.625 & -0.625 & -0.625 & -0.625 & -0.625 & -0.625 \\
        $\beta_{2VX}$ & 0 & 0 & 0 & 0 & 0 & 0 & 0 & \textbf{-1.5} \\
        $\beta_{2U}$ & 1 & 1 & 1 & 1 & 1 & 1 & 1 &  1 \\
        $\tau_1$ & -1.1 & -1.1 & -1.1 & -1.1 & -1.1 & -1.1 & -1.1 &  -1.1 \\
        $\tau_2$ & -0.6 & -0.6 & -0.6 & -0.6 & -0.6 & -0.6 & -0.6 &  -0.6 \\
        $\tau_{1V}$ & 0 & 0 & 0 & 0 & 0 & \textbf{-0.25} & \textbf{-0.25} & 0 \\
        $\tau_{2V}$ & 0 & 0 & 0 & 0 & 0 & 0 & \textbf{-0.25} & 0 \\
        $\tau_X$ & 0.25 & 0.25 & 0.25 & 0.25 & 0.25 & 0.25 & 0.25 & 0.25 \\
        $\tau_U$ & 0.25 & 0.25 & 0.25 & 0.25 & 0.25 & 0.25 & 0.25 & 0.25 \\
        $\tau_{2U}$ & 0 & 0  & 0  & 0 & \textbf{-2}  & 0  & 0   & 0  \\
        \bottomrule
    \end{tabular}
\end{table}

\begin{table}[p]
    \centering
    \caption{Estimation methods for simulation study.}\label{tab:methods}
    \begin{tabular}{llp{3in}}
        \toprule
        Name & Estimator & Methods \\
        \midrule
        TND, logit & $\widehat{\Psi}_{om}^*(X)$ & conventional approach to TND based on coefficient from correctly-specified logistic regression of $Y^*$ on $(1, V, X)'$. \\
        \addlinespace[1em]
        TND, om & $\widehat{\Psi}^*_{om}$ & Proposed estimator in expression \ref{eqn:om_estimator}. Nuisance terms $\mu_v^*(X)$ estimated via logistic regression of $Y^*$ on $(1, V, X, VX)'$.  \\
        \addlinespace[1em]
        TND, ipw & $\widehat{\Psi}^*_{ipw}$ & Proposed estimator in expression \ref{eqn:ipw_estimator}. Nuisance term $\pi_0^*(X)$ estimated via logistic regression of $V$ on $(1, X)'$ among those with $Y^*=0$. \\
        \addlinespace[1em]
        TND, dr & $\widehat{\Psi}^*_{dr}$ & Proposed estimator in expression \ref{eqn:dr_estimator}. Nuisance terms $\mu_v^*(X)$ and $\pi_0^*(X)$ estimated as above. \\
        \addlinespace[1em]
        DiD & $\widehat{\Psi}_{om}$ &  Alternative estimator introduced in expression \ref{eqn:om_estimator_cohort}. Nuisance terms $\mu_v(X)$ estimated via logistic regression of $Y$ on $(1, V, X, VX)'$ in the full sample. \\
        \addlinespace[1em]
        cohort, U unmeasured & $\widehat{\Psi}_{cohort}$ & Standardization estimator in expression \ref{eqn:cohort_estimator} assuming $U$ is unavailable. Nuisance terms $\mu_v(X)$ estimated via logistic regression of $Y$ on $(1, V, X, VX)'$ in the full sample. \\ 
        \addlinespace[1em]
        cohort, U measured & $\widehat{\Psi}_{cohort,U}$ & Standardization estimator in expression \ref{eqn:cohort_u_estimator} assuming $U$ is available. Nuisance terms $\mu_v(X,U)$ estimated via logistic regression of $Y$ on $(1, V, X, VX, U)'$ in the full sample. \\ 
        \bottomrule
    \end{tabular}
\end{table}

\newcommand*{\TableHead}[1]{\multicolumn{1}{p{3em}}{\centering\hskip0pt#1}}
\newcommand*{\TableHeadd}[1]{\multicolumn{1}{p{4em}}{\centering\hskip0pt#1}}
\newcommand*{\TableHeaddd}[1]{\multicolumn{1}{p{7em}}{\centering\hskip0pt#1}}

\begin{table}
\caption{Results of simulation study: 1000 simulations of sample size $n=$15,000 across 9 scenarios with TND selection probabilities between 9\% and 15\%. Bias is estimated mean bias, SE is estimated Monte Carlo standard error, Coverage is estimated coverage probability of 95\% CI}\label{tab:sims}
\centering
\begin{tabular}{lddddddd}
\toprule
Statistic & \TableHeadd{$\widehat{\Psi}_{om}^*(X)$}  & \TableHeadd{$\widehat{\Psi}_{om}^*$}  & \TableHeadd{$\widehat{\Psi}_{ipw}^*$}  & \TableHeadd{$\widehat{\Psi}_{dr}^*$} & \TableHeadd{$\widehat{\Psi}_{om}$} & \TableHeadd{$\widehat{\Psi}_{cohort,U}$} & \TableHeadd{$\widehat{\Psi}_{cohort}$} \\
\midrule
\addlinespace[0.3em]
\multicolumn{8}{l}{\textit{scenario 1: no unmeasured confounding}}\\
\hspace{1em}Bias & 0.004 & 0.004 & 0.004 & 0.004 & 0.004 & 0.002 & 0.002\\
\hspace{1em}SE & 0.051 & 0.051 & 0.051 & 0.051 & 0.051 & 0.040 & 0.040\\
\hspace{1em}Coverage & 0.945 & 0.944 & 0.943 & 0.944 & 0.944 & 0.947 & 0.950\\
\addlinespace[0.3em]
\multicolumn{8}{l}{\textit{scenario 2: equi-confounding}}\\
\hspace{1em}Bias & 0.000 & 0.000 & 0.000 & 0.000 & 0.001 & 0.000 & 0.078\\
\hspace{1em}SE & 0.036 & 0.036 & 0.036 & 0.036 & 0.036 & 0.029 & 0.085\\
\hspace{1em}Coverage & 0.964 & 0.957 & 0.958 & 0.957 & 0.965 & 0.949 & 0.308\\
\addlinespace[0.3em]
\multicolumn{8}{l}{\textit{scenario 3: direct effect of V on I = 1}}\\
\hspace{1em}Bias & 0.040 & 0.039 & 0.039 & 0.039 & 0.040 & -0.001 & 0.077\\
\hspace{1em}SE & 0.057 & 0.057 & 0.057 & 0.057 & 0.058 & 0.030 & 0.085\\
\hspace{1em}Coverage & 0.838 & 0.844 & 0.844 & 0.843 & 0.839 & 0.943 & 0.342\\
\addlinespace[0.3em]
\multicolumn{8}{l}{\textit{scenario 4: equi-confounding violated}}\\
\hspace{1em}Bias & 0.047 & 0.046 & 0.046 & 0.046 & 0.047 & 0.001 & 0.079\\
\hspace{1em}SE & 0.066 & 0.066 & 0.066 & 0.066 & 0.066 & 0.029 & 0.087\\
\hspace{1em}Coverage & 0.831 & 0.836 & 0.837 & 0.836 & 0.833 & 0.955 & 0.308\\
\addlinespace[0.3em]
\multicolumn{8}{l}{\textit{scenario 5: equi-selection violated}}\\
\hspace{1em}Bias & -0.097 & -0.097 & -0.097 & -0.097 & -0.097 & 0.001 & -0.040\\
\hspace{1em}SE & 0.105 & 0.106 & 0.106 & 0.106 & 0.105 & 0.052 & 0.061\\
\hspace{1em}Coverage & 0.476 & 0.474 & 0.479 & 0.479 & 0.477 & 0.954 & 0.880\\
\addlinespace[0.3em]
\multicolumn{8}{l}{\textit{scenario 6: equal effect of V on T}}\\
\hspace{1em}Bias & 0.000 & -0.001 & -0.001 & -0.001 & 0.000 & -0.082 & -0.021\\
\hspace{1em}SE & 0.039 & 0.040 & 0.040 & 0.040 & 0.039 & 0.086 & 0.036\\
\hspace{1em}Coverage & 0.957 & 0.957 & 0.957 & 0.958 & 0.959 & 0.151 & 0.909\\
\addlinespace[0.3em]
\multicolumn{8}{l}{\textit{scenario 7: unequal effect of V on T}}\\
\hspace{1em}Bias & 0.106 & 0.105 & 0.105 & 0.105 & 0.106 & 0.000 & 0.078\\
\hspace{1em}SE & 0.117 & 0.116 & 0.116 & 0.116 & 0.117 & 0.030 & 0.086\\
\hspace{1em}Coverage & 0.339 & 0.348 & 0.350 & 0.350 & 0.332 & 0.951 & 0.314\\
\addlinespace[0.3em]
\multicolumn{8}{l}{\textit{scenario 8: effect heterogeneity}}\\
\hspace{1em}Bias & -0.042 & -0.002 & -0.002 & -0.002 & -0.039 & -0.001 & 0.226\\
\hspace{1em}SE & 0.097 & 0.097 & 0.097 & 0.097 & 0.095 & 0.074 & 0.073\\
\hspace{1em}Coverage & 0.923 & 0.944 & 0.944 & 0.944 & 0.923 & 0.938 & 0.133\\
\bottomrule
\end{tabular}
\end{table}

\begin{table}[p]
    \centering
    \caption{Robustness of estimators to misspecification of nuisance functions: 1000 simulations of sample size $n=$15,000 from modified processes based on scenario 8 in eTable 3. Bias is estimated mean bias, SE is estimated monte carlo standard error, Coverage is estimated coverage probability of 95\% CI}\label{tab:sims2}
    \begin{tabular}{lddddddd}
    \toprule
    Statistic & \TableHeadd{$\widehat{\Psi}_{om}^*(X)$}  & \TableHeadd{$\widehat{\Psi}_{om}^*$}  & \TableHeadd{$\widehat{\Psi}_{ipw}^*$}  & \TableHeadd{$\widehat{\Psi}_{dr}^*$} & \TableHeadd{$\widehat{\Psi}_{om}$} & \TableHeadd{$\widehat{\Psi}_{cohort,U}$} & \TableHeadd{$\widehat{\Psi}_{cohort}$} \\
    \midrule
    \addlinespace[0.3em]
\multicolumn{8}{l}{\textit{scenario 8: both correctly specified}}\\
\hspace{1em}Bias & -0.042 & -0.002 & -0.002 & -0.002 & -0.039 & -0.001 & 0.226\\
\hspace{1em}SE & 0.097 & 0.097 & 0.097 & 0.097 & 0.095 & 0.074 & 0.073\\
\hspace{1em}Coverage & 0.923 & 0.944 & 0.944 & 0.944 & 0.923 & 0.938 & 0.133\\
\addlinespace[0.3em]
\multicolumn{8}{l}{\textit{scenario 8: propensity score model misspecified}}\\
\hspace{1em}Bias & -0.037 & -0.001 & -0.012 & -0.001 & -0.038 & 0.000 & 0.215\\
\hspace{1em}SE & 0.102 & 0.101 & 0.101 & 0.101 & 0.098 & 0.076 & 0.075\\
\hspace{1em}Coverage & 0.941 & 0.954 & 0.954 & 0.954 & 0.942 & 0.953 & 0.198\\
\addlinespace[0.3em]
\multicolumn{8}{l}{\textit{scenario 8: outcome model misspecified}}\\
\hspace{1em}Bias & -0.062 & 0.022 & -0.004 & 0.004 & -0.023 & 0.028 & 0.256\\
\hspace{1em}SE & 0.093 & 0.093 & 0.095 & 0.094 & 0.090 & 0.071 & 0.069\\
\hspace{1em}Coverage & 0.911 & 0.949 & 0.957 & 0.958 & 0.944 & 0.942 & 0.052\\
\addlinespace[0.3em]
\multicolumn{8}{l}{\textit{scenario 8: both misspecified}}\\
\hspace{1em}Bias & -0.097 & 0.018 & -0.091 & -0.023 & -0.046 & 0.025 & 0.285\\
\hspace{1em}SE & 0.096 & 0.096 & 0.113 & 0.099 & 0.090 & 0.075 & 0.070\\
\hspace{1em}Coverage & 0.829 & 0.947 & 0.894 & 0.950 & 0.930 & 0.933 & 0.028\\
\bottomrule
\end{tabular}
\end{table}
\newpage
\clearpage
% $\varphi$
 \printbibliography[heading=subbibliography]
\end{refsection}
\end{appendix}

%TC:endignore

\end{document}